\documentclass[%
 reprint,
 amsmath,amssymb,
 aps,
]{revtex4-2}

\usepackage{graphicx}
\usepackage{dcolumn}
\usepackage{bm}
\usepackage{hyperref}

\usepackage{braket}
\usepackage{xcolor}

\begin{document}

\preprint{APS/123-QED}

\title{The use of Peres lattices in periodically driven systems}

\author{Lukáš Honsa}
 \email{lukas.honsa@matfyz.cuni.cz}
 \author{Jan Střeleček}
 \author{Jakub Novotný}
\author{Pavel Cejnar}
\email{pavel.cejnar@matfyz.cuni.cz}
\affiliation{%
 Institute of Particle and Nuclear Physics, Faculty of Mathematics and Physics,
 Charles University, V Holešovičkách 2, 18000 Prague, Czech Republic
}%




\date{\today}

\begin{abstract}
We demonstrate the strength of the method of Peres lattices in periodically driven quantum systems.
The method, which has previously been used mostly in stationary systems, enables us to efficiently detect resonances in the driven system, to monitor the onset of chaos, and to recognize critical properties of the Floquet modes.
It also allows quick comparisons of the spectra of Floquet modes for various driving Hamiltonians and transparent tests of the iterative approximation techniques based on effective stationary Hamiltonians.
\end{abstract}

\maketitle


\section{Introduction}

Discrete spectra of stationary states contain complete information on the dynamics of closed quantum systems.
Suitable techniques for the display of joint spectral properties for large amounts of stationary states may therefore substantially help to reveal and better understand the essential local and global features of such systems, particularly features related to integrability or chaos \cite{Reichl2004, Haake2001, Gutzwiller1990}, and to quantum phase transitions in the ground and excited states \cite{Carr2010, Sachdev2011, Heyl2018, Cejnar2021}.
In this context, the method of Peres lattices \cite{Peres1984a} (also called quantum webs \cite{Reichl2004}) proved to be a very efficient tool.
It depicts the spectrum of a finite bound quantum system with a limited number of degrees of freedom $f$ (typically $f=2$) as a mesh of points corresponding to individual eigenstates in the $f$-dimensional space (typically a~plane) whose perpendicular axes represent expectation values of some relevant observables.
The patterns observed in this scatter plot carry a message that can be used in various contexts. 

The method was originally proposed by Asher Peres in 1984 for analyses of quantum chaos, in particular in systems with $f=2$, where the Peres lattice can be seen as a quantum analog of the classical Poincaré surface of section \cite{Peres1984a}.
The lattice enables one to visually distinguish the regular and chaotic domains of the spectrum according to ordered and disordered distributions of the mesh points, respectively \cite{Ramirez2025,Villaseor2022,Lobez2021,Reichl2018,Barr2017,Porter2017,Magnani2017,Relano2016,Magnani2016,Robles2015,Basta2014,Macek2014,Leviatan2012,Macek2011,Stransky2009,Stepanov2008,Ree1999,Camargo1994,Srivastava1990b}.
The Peres method has also proven to be very helpful in the location of excited-state quantum phase transitions \cite{Corps2021,Macek2019,Zhu2019,Kloc2018,Ramos2017,Lobez2016,Stransky2015,Magnani2014} and in the identification of related quantum phases in the spectrum \cite{Cejnar2021,Rodr2018,Kloc2017a,Kloc2017b}.
It also facilitates the recognition of some eigenstates with anomalous properties, e.g., non-ergodic or scarred states \cite{Lerose2025,Pilat2021a,Pilat2021b}, and the study of quantum-classical correspondence \cite{Robb1998,Srivastava1990a,Srivastava1990c,Feingold1985}.

This paper explores the applicability of the Peres lattice method to periodically driven quantum systems. 
Such systems have long been used as theoretical tools for studying the order-to-chaos transition in both the classical and quantum realms \cite{Reichl2004,Haake2001}. 
Currently, they are experiencing an intense revival due to their rapidly improving experimental accessibility \cite{Sieberer2019, Tomkovic2017,Gadway2013,Chaudhury2009, Grifoni1998} and their great potential in quantum simulations \cite{Eckstein2024,Anand2024,Mizuta2023,Olsacher2022, Santhanam2022,Sieberer2019,Kyriienko2018}. It also turned out that periodic driving of a closed quantum system can induce critical effects quite analogous to quantum phase transitions in stationary systems \cite{Garcia-Mata2021,Bandyopadhyay2015,Bastidas2014a, Bastidas2014b,Engelhardt2013}, generate exotic dynamical states of matter called time crystals \cite{Zaletel2023,Sacha2018}, and lead to fractal butterfly Floquet spectra \cite{Sharma2018, Bandyopadhyay2010, Wang2010}.

In the context of periodic driving, the Peres lattice method must undergo certain modifications.
First, the stationary energy eigenstates, which are no longer defined, must be replaced by the eigenstates of the evolution operator over one period, i.e., by the so-called Floquet modes of the time-periodic Hamiltonian.
Second, it is known that a classical system with $f$ degrees of freedom with a periodic time-dependent Hamiltonian can be cast as a system with ${f+1}$ degrees of freedom with a time-independent Hamiltonian.
This allows for chaotic behavior even in an ${f=1}$ driven system, whose classical return map after one period becomes fully analogous to an ${f=2}$ Poincaré surface of section.
Therefore, quantum Peres lattices of a periodically driven system with $f$ degrees of freedom need to be drawn with the aid of ${f+1}$ observables. 

In this paper, the method of Peres lattices is applied to a periodically driven ${f=1}$ system of the Lipkin--Meshkov--Glick type \cite{Lipkin1965188} (it describes a fully connected spin or qubit system).
The unperturbed Hamiltonian exhibits a continuous ground-state quantum phase transition, and above the critical point also an excited-state quantum phase transition \cite{Cejnar2021}. 
We set the parameters so that the periodic driving term of the full Hamiltonian probes both stationary phases of the original system.
The driving term takes different time dependencies and operator forms, but in most cases it represents only a weak perturbation of the stationary system.
We also analyze to what extent the exact evolution with the full time-dependent Hamiltonian can be represented by an evolution with an effective stationary Hamiltonian (so-called Floquet Hamiltonian) written as a series in powers of the driving period and strength \cite{Lando2020, Kuwahara2016, Bandyopadhyay2015, Bastidas2014a, Bastidas2014b, Goldman2014, Rahav2003, Scharf1988a, Scharf1988b}.
We demonstrate that the Peres lattice method yields quick insights into all the addressed problems.

The plan of the paper is as follows:
In Sect.\,\ref{sec:ham} we specify alternative forms of the time-dependent Hamiltonian.
In Sect.\,\ref{sec:LM} we describe basic properties of the stationary system. 
In Sect.\,\ref{sec:kick} we present Peres lattices obtained under various conditions and for various observables in the case of impulse (delta kicked) periodic driving. We demonstrate general features related to the onset of chaos and the survival of quantum criticality in the driven system and discuss the approximation through truncated Floquet Hamiltonians. 
In Sect.\,\ref{sec:VD} we present similar results for other forms of periodic driving. 
In Sect.\,\ref{sec:conc} we summarize results and draw conclusions.
Appendices~\ref{sec:App_T_0} and \ref{sec:App_Convergence} describe some more technical issues and Appendix~\ref{sec:App_num} comments on the numerical methods.
For the sake of simplicity, all quantities considered below are taken as dimensionless.

\section{Hamiltonians\label{sec:ham}}

We examine periodically driven systems, i.e., systems with Hamiltonian $\widehat{H}(t) = \widehat{H}(t +T)$, where $T$ is the time period. We set the following form of the Hamiltonian:
\begin{eqnarray}
	\widehat{H} (t) = \widehat{H}_0 + \eta g(t) \widehat{H}^\prime, 
    \label{eq:DrivenLH}
\end{eqnarray}
where $\widehat{H}_0$ is a stationary (unperturbed) Hamiltonian, $\widehat{H}^\prime$ is a driving operator, $\eta\geq 0$ is a driving strength, and $g (t) = g(t+T)$ is a periodic driving function with zero average, $\int_0^T g(t) dt = 0$, and with the integrated absolute value satisfying a condition $\int_0^T |g(t)| dt \sim 1$.

In this work, we discuss Hamiltonians composed of quasispin (or angular momentum) operators $\widehat{J}_k$ with ${k \in \{x, y ,z\}}$ acting in the Hilbert space associated with a single ${(2j+1)}$-dimensional eigenspace of~$\widehat{J}^2$. 
The stationary system corresponds to the Lipkin--Meshkov--Glick model \cite{Lipkin1965188} with Hamiltonian
\begin{eqnarray}
	\widehat{H}_0 = \widehat{J}_z - \frac{\kappa}{2j}\widehat{J}_x^2 
    \label{eq:LH}.
\end{eqnarray}  
This can be seen as a fully connected system of $2j$ qubits in the subspace of fully exchange-symmetric states, with
\begin{equation}
\widehat{J}_k = \frac{1}{2}\sum_{s = 1}^{2j} \sigma_k^{(s)},
\end{equation}
where $\sigma_k^{(s)}$ denote Pauli matrices of the qubit on site $s$. 
Consequently, the operators ${\widehat{J}_z\propto\sum_{s}\sigma_z^{(s)}}$ and ${\widehat{J}_x^2\propto\sum_{s,s'}\sigma_x^{(s)}\sigma_x^{(s')}}$ in Hamiltonian~\eqref{eq:LH} represent, respectively, the self-energies of individual qubits and their mutual interactions.
The parameter $\kappa$ sets the proportion of both these terms and controls the critical properties of the stationary system. 
We briefly discuss basic features of this model in the next section.

\begin{figure}
	\includegraphics[width=\columnwidth]{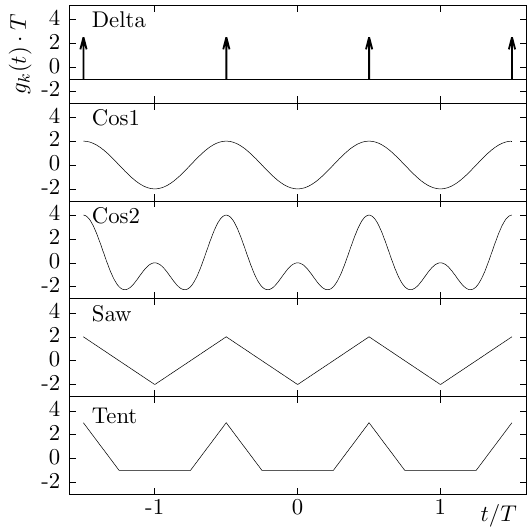}
	\caption{\label{fig:gs}
    Various driving functions $g (t)$ from Eqs.\,\eqref{eq:g_Delta}--\eqref{eq:g_Tent}. The black vertical arrows represent the delta functions.}
\end{figure}

We will employ the driving functions defined as follows:
	\begin{subequations}
		\newsavebox{\mycasesB}
		\begin{align}
			\sbox{\mycasesB}{$\displaystyle \hspace{-2mm}g(t) = \left\{\vphantom{\begin{array}{@{}c@{}}{}
						~\widehat{J}^2\\
						~\widehat{J}^2\\ 
						~\\
						~\\
						~\\~\\~\\~\\~\\~\\~\\~\\~\\~\\~\\
				\end{array}}\right.$}
			\raisebox{-.87\ht\mycasesB}[0pt][0pt]{\usebox{\mycasesB}}
			&-\frac{1}{T}+\delta(t-T/2),                &\hspace{-10mm}\rm{Delta},                                                              \label{eq:g_Delta} \\
			&-\frac{2}{T} \cos{\frac{2 \pi t}{T}}, &\rm{Cos1},                                                               \label{eq:g_Cos1}\\
			&\frac{2}{T} \left( - \cos{\frac{2 \pi t}{T}} +\cos{\frac{4 \pi t}{T}} \right), &\rm{Cos2},   \label{eq:g_Cos2}\\
			&-\frac{2}{T} + \frac{8t}{T^2}\chi_{(0, T/2]}(t)\nonumber\\&~~~~~+ \left(\frac{8}{T} - \frac{8t}{T^2}\right)\chi_{(T/2, T]}(t),  &\rm{Saw}                                                                                                                               \label{eq:g_Saw},\\
			&-\frac{1}{T} + \left(-\frac{4}{T} + \frac{16t}{T^2}\right)\chi_{(T/4, T/2]}(t) \nonumber\\&~~~~~+\left(\frac{12}{T} - \frac{16t}{T^2}\right)\chi_{(T/2, 3T/4]}(t), &\rm{Tent},                                                        \label{eq:g_Tent}
		\end{align}
	\end{subequations}
where $t\in [0, T]$. These functions are extended to any $t \in \mathbb{R}$ using $g(t) = g(t + n T)$ with $n \in \mathbb{Z}$. The symbol $\delta(x)$ denotes the Dirac delta function and $\chi_{\left(t_1, t_2\right]}(t)$ stands for the characteristic interval function defined by $\chi_{\left(t_1, t_2\right]} (t) = 1$ for $t \in (t_1, t_2]$ and $\chi_{\left(t_1, t_2\right]} (t) = 0$ otherwise. 
The different driving functions in Eqs.\,\eqref{eq:g_Delta}--\eqref{eq:g_Tent} are plotted in Fig. \ref{fig:gs}.
Finally, we will employ the following forms of the driving operator:
\begin{subequations}
	\newsavebox{\mycases}
	\begin{align}
		\sbox{\mycases}{$\displaystyle \widehat{H}^\prime=\left\{\vphantom{\begin{array}{@{}c@{}}{}
					\widehat{J}_z,\\
					\widehat{J}_z^2/j,,\\ 
					\widehat{J}_x.\\~
			\end{array}}\right.$}
		\raisebox{-.58\ht\mycases}[0pt][0pt]{\usebox{\mycases}}
		\widehat{J}_z,\!      \label{eq:Hprime_Jz} \\
		\widehat{J}_z^2/j,\!\!\!\!\!\!\!      \label{eq:Hprime_squareJz}\\
		\widehat{J}_x.\!  \label{eq:Hprime_Jx}
	\end{align}
\end{subequations}

For all choices of $g(t)$ and $\widehat{H}^\prime$, the total squared angular momentum ${\widehat{J}^2=\widehat{J}_x^2 + \widehat{J}_y^2 + \widehat{J}_z^2}$ is conserved during the evolution. Consequently, the evolution can be confined to the $(2j+1)$-dimensional eigenspace of $\widehat{J}^2$ with the eigenvalue $j(j+1)$, where the quantum number $j$ is a~fixed parameter that determines the size of the system.
The classical limit of the system is achieved for an infinite size, i.e., $j \rightarrow \infty$. 
In that limit, operators can be replaced by ordinary variables:
\begin{eqnarray}
	\mathcal{J}_k  &=& \lim_{j \rightarrow \infty} \frac{\widehat{J}_k}{j},~k\in\{x, y, z\}, \label{eq:classica_J} \\
	\mathcal{H}(t) &=& \lim_{j \rightarrow \infty} \frac{\widehat{H}(t)}{j}, \label{eq:Classical_limit} \\
	       &=&  \mathcal{J}_z - \frac{\kappa}{2} \mathcal{J}_x ^2 + \eta g (t)  \mathcal{J}_k. \label{eq:Class_H}
\end{eqnarray}
The classical phase space is the Bloch sphere
\begin{equation}\label{eq:Bloch}
(\mathcal{J}_x,\mathcal{J}_y,\mathcal{J}_z)=(\sin\vartheta\cos\varphi,\sin\vartheta\sin\varphi,\cos\vartheta)
\end{equation}
with unit radius $|\vec{\mathcal{J}}| = 1$, on which the longitude $\varphi$ represents a coordinate and the projection $\cos\vartheta$ the associated momentum.

The classical time-dependent Hamiltonian $\mathcal{H}(\vartheta,\varphi,t)$ with period $T$ defined in this $f=1$ phase space, Eq.\,\eqref{eq:Class_H}, can be cast as a stationary Hamiltonian of an extended system with $f=2$ degrees of freedom:
\begin{equation}\label{eq:Hex}
    \mathcal{H}_{\rm ext}(\vartheta,\varphi,\phi,\mathcal{I})=\mathcal{H}\left(\vartheta,\varphi,\frac{T}{2\pi}\phi\right)+\frac{2\pi}{T}\mathcal{I},
\end{equation}
where $\phi$ is an additional angular coordinate and $\mathcal{I}$ the associated momentum \cite{Reichl2004}.
With this Hamiltonian, the old phase-space variables move exactly as in the original driven system, while the evolution of the new coordinate $\phi$ emulates the monotonously running time and the momentum $\mathcal{I}$ changes so that the overall energy \eqref{eq:Hex} is conserved.

\section{Stationary system\label{sec:LM}}

In this section, we review basic features of the stationary quantum system $\widehat{H}_0$, namely its critical properties and classical-limit dynamics, and present the unperturbed forms of various Peres lattices.

\subsection{Quantum phase transitions\label{sec:QPT}}

\begin{figure}
\includegraphics[width=\columnwidth]{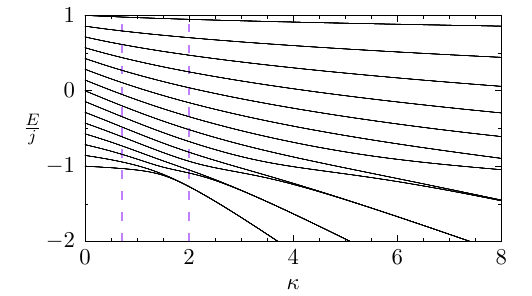}
\caption{\label{fig:LH}Spectrum of the stationary Hamiltonian $\widehat{H}_0$ from Eq.\,\eqref{eq:LH} for ${j = 7}$. The parameter values used in the following text are indicated by dashed vertical lines.}
\end{figure}

As indicated above, the stationary system is a fully connected system of qubits with the Hamiltonian defined by Eq.\,\eqref{eq:LH}. 
The energy spectrum for a moderate value of $j$ is shown in Fig.\,\ref{fig:LH}. 
In the limit $j\to\infty$, the model exhibits a second-order quantum phase transition between the non-interacting and interacting ground-state phases, which occurs at the parameter value $\kappa = \kappa_c = 1$.
While for $\kappa < \kappa_c$ the ground state is determined only by the term $\widehat{J}_z$ (it coincides with the spin-down state of all qubits), for $\kappa > \kappa_c$ it results from an interplay of both $\widehat{J}_z$ and $\widehat{J}_x^2$ terms (it is an entangled superposition of spin-up and down states of the qubits).
For $\kappa > \kappa_c$, the system exhibits an excited-state quantum phase transition at the critical energy $E=E_c = -j$.
Finite-size precursors of these phase transitions are seen in Fig.\,\ref{fig:LH}.
In our simulations of driven dynamics, we will use two values of $\kappa$, namely $\kappa = 0.7$ and $\kappa = 2.0$, which are below and above the critical point. 

The ground-state quantum phase transition \cite{Lipkin1965188, Gilmore1978} at $\kappa=\kappa_c$ and the subsequent excited-state quantum phase transition \cite{Santos2016, Engelhardt2015, Ribeiro2008} at $E=E_c$ can be understood in terms of spontaneous breaking of parity.
Indeed, the Hamiltonian \eqref{eq:LH} can be shown to conserve the parity defined~as 
\begin{equation}\label{eq:parit}
\hat{\Pi}=(-1)^{\widehat{J}_z+j}
\end{equation}
with eigenvalues $\pm 1$.
While all Hamiltonian eigenstates for $\kappa<\kappa_c$ and the eigenstates with $E>E_c$ for $\kappa>\kappa_c$ have well-defined parity quantum numbers (alternating positive and negative values as the energy increases), the eigenstates with $E<E_c$ for $\kappa>\kappa_c$ converge to degenerate parity doublets in the limit $j\to\infty$, constituting the parity-breaking phase.
For finite values of $j$, the degeneracy in the broken phase is not exact, but the corresponding pairs of positive- and negative-parity levels are very close to each other even for moderate $j$, see Fig.\,\ref{fig:LH}.

\subsection{Classical dynamics}

\begin{figure}
	\includegraphics[width=\columnwidth]{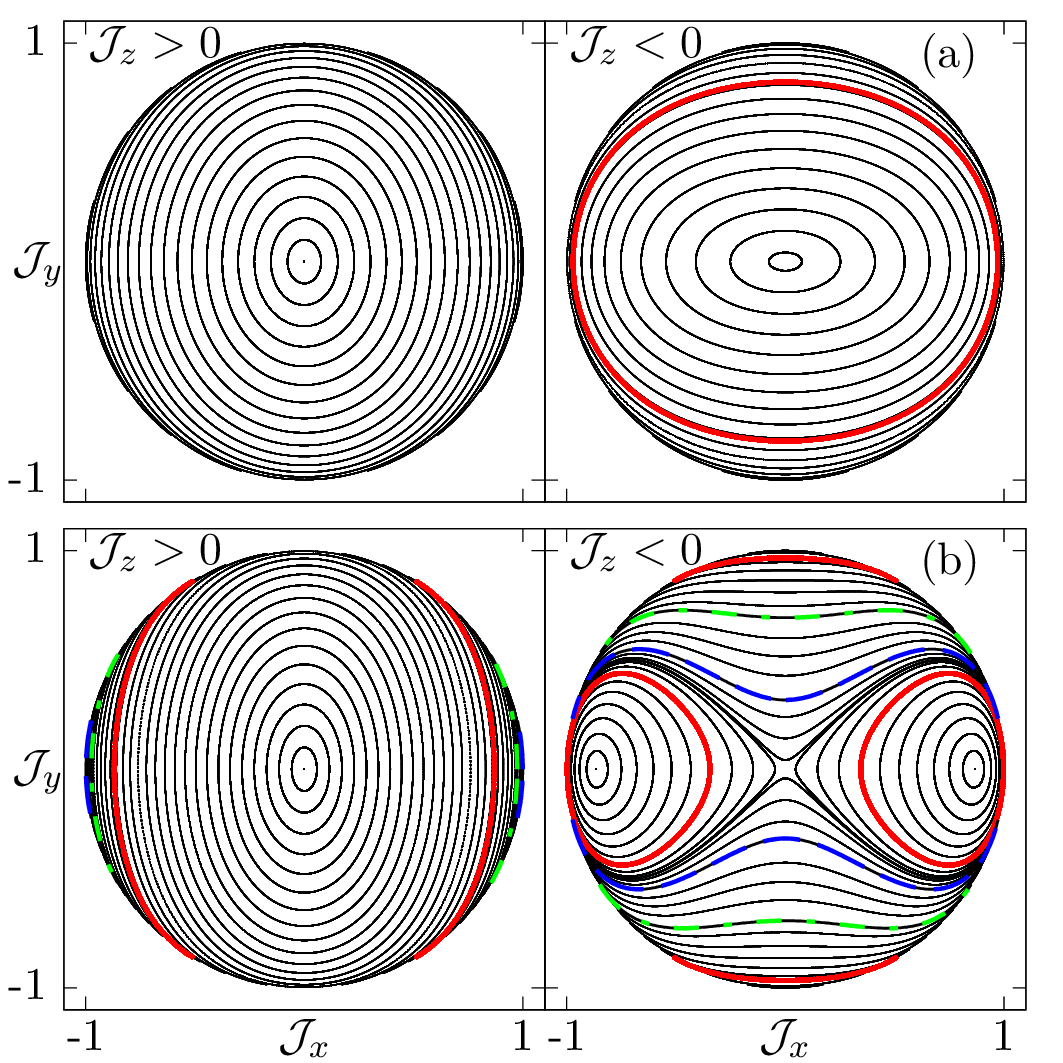}
	\caption{\label{fig:LH_CL_PR}
    Classical phase space trajectories of the stationary system for (a) $\kappa = 0.7$ (upper plots) and (b) $\kappa = 2$ (lower plots). The left and right plots, respectively, show the upper and lower hemispheres of the Bloch sphere. Trajectories with some specific time periods $\tau$, which will be important in later discussions, are highlighted.}
\end{figure}

The classical unperturbed Hamiltonian from Eq.\,\eqref{eq:Class_H} is
\begin{eqnarray}
	\mathcal{H}_0 = \mathcal{J}_z -\frac{\kappa}{2} \mathcal{J}_x^2,  \label{eq:LHC}
\end{eqnarray}
the corresponding phase space being associated with the $|\vec{J}| = 1$ Bloch sphere.   
Classical trajectories, which coincide with energy contours of the function~\eqref{eq:LHC}, are depicted, for the above-indicated two values of $\kappa$, in Fig.\,\ref{fig:LH_CL_PR}.
We see that for $\kappa>\kappa_c$ the phase space divides into two separate domains, which are connected to the parity-conserving and parity-breaking phases of the Hamiltonian. 
While the parity-conserving phase is associated with trajectories ranging around the whole Bloch sphere, the parity-breaking phase corresponds to the trajectories confined in two symmetrically placed ${\mathcal{J}_x<0}$ and ${\mathcal{J}_x>0}$ phase-space regions (lobes), see Fig.\,\ref{fig:LH_CL_PR}(b).

\begin{figure}
    \includegraphics[width=\columnwidth]{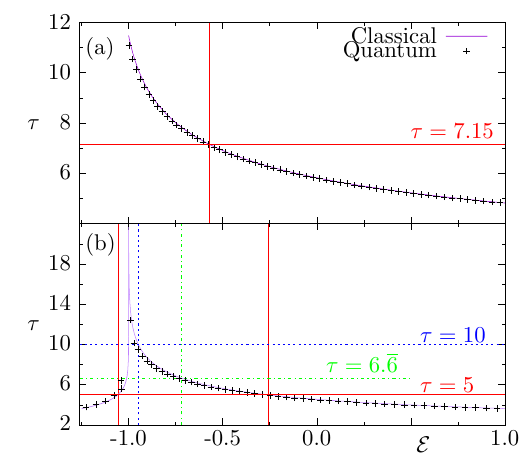}
    \caption{\label{fig:Etau} 
    The period $\tau$ of classical trajectories for Hamiltonian \eqref{eq:LHC} with (a) $\kappa = 0.7$ and (b) $\kappa = 2$ as a function of scaled energy $\mathcal{E}=E/j$. The purple line depicts the classical calculation of $\tau(\mathcal{E})$, the black crosses show the estimate \eqref{eq:pairs} from the spacings of quantum energy levels for $j=30$. Energies and periods corresponding to the trajectories highlighted in Fig.\,\ref{fig:LH_CL_PR} are marked.}
\end{figure}

The dependence of the period $\tau$ of a given trajectory on the classical energy $\mathcal{E} = \mathcal{H}(\vec{\mathcal{J}})$ with the scaled Hamiltonian \eqref{eq:Classical_limit} is shown in Fig.\,\ref{fig:Etau}, again for the two values of parameter $\kappa$. 
We also plot the points 
\begin{equation}
  (\mathcal{E}_i, \tau_i)=\left\{\begin{array}{ll}
  (\frac{E_{i+1}+E_i}{2j},\frac{2\pi}{E_{i+1} - E_i}) & {\rm for\ }(\kappa,E_{i})\notin{\rm BP,} \\
  (\frac{E_{i+2}+E_i}{2j},\frac{2\pi}{E_{i+2} - E_i}) & {\rm for\ }(\kappa,E_{i+2})\in{\rm BP,}
  \end{array}
  \right.
  \label{eq:pairs}
\end{equation}
where $E_i$ are the eigenvalues of the stationary Hamiltonian satisfying $\widehat{H}_0 \ket{E_i}= E_i \ket{E_i}$, and BP stands for the parity-breaking phase defined as the domain with $\kappa>\kappa_c$ and $E<E_c$.
The validity of the approximation $\tau(\mathcal{E}_i)\approx\tau_i$ for large values of~$j$ follows from the semiclassical formula $\Delta E=2\pi\hbar/\tau$ for the spacing $\Delta E$ between energy levels of $f=1$ systems, which applies separately in each simply connected phase space domain associated with a given energy interval.
In the parity-breaking phase, the evaluation of periods for the trajectories located in both symmetric phase space domains requires using the second adjacent level spacing, which avoids problems with nearly degenerate parity doublets, cf.\,Fig.\,\ref{fig:LH}.
Let us note that in our model, the effective Planck constant is taken as $j^{-1}$.

In Fig.\,\ref{fig:Etau}(b) we see that the period $\tau(\mathcal{E})$ of the $\kappa=2$ system diverges at $\mathcal{E}=-1$.
This is a consequence of the infinite-period separatrix trajectory passing through the hyperbolic stationary point at $(\mathcal{J}_x,\mathcal{J}_y,\mathcal{J}_z)=(0,0,-1)$, which is a hallmark of the $\kappa > \kappa_c$ regime of dynamics; cf.\,Fig.\,\ref{fig:LH_CL_PR}(b).
This classical stationary point corresponds to the excited-state quantum phase transition between the parity-breaking and parity-conserving phases at ${E=E_c}$ \cite{Cejnar2021, Engelhardt2015}.

\subsection{Peres lattices}

Since our unperturbed model has only one degree of freedom, the conservation of energy $\widehat{H}_0$ implies that it is integrable.
Therefore, we anticipate that individual points in the scatter plot of the expectation values $\bra{E_i}\widehat{O}\ket{E_i}\equiv\langle\widehat{O}\rangle_i$ versus $\bra{E_i}\widehat{H}_0\ket{E_i}\equiv\langle\widehat{H}_0\rangle_i\equiv E_i$, where $\widehat{O}$ is an arbitrary observable and $\ket{E_i}$ are eigenvectors of $\widehat{H}_0$, fill in curves defining a more or less smooth functional dependence.
This follows from the classical limit, when $\widehat{O}$ becomes a function $\mathcal{O}(\vartheta,\varphi)$ in the phase space (Bloch sphere) and the expectation values $\langle\widehat{O}\rangle_i$ are replaced by asymptotic time averages $\langle\mathcal{O}\rangle$ of the classical quantity $\mathcal{O}$ over individual trajectories.
These averages define a new conserved quantity $\overline{\mathcal{O}}(\vartheta,\varphi)$ in the phase space \cite{Peres1984a}, which in the $f=1$ system must be a~function of the other conserved quantity $\mathcal{H}_0(\vartheta,\phi)$.
Even the Peres lattice of expectation values $\langle\widehat{O}_1\rangle_i$ and $\langle\widehat{O}_2\rangle_i$ of any other observables will typically contain just one-dimensional chains of points. 
Thus, in the unperturbed model, the Peres concept ceases to show its strength.
Nevertheless, we present here a few examples of the Peres lattice of the stationary system for further reference, when lattices of the same type for the system with a periodic perturbation will become nontrivial due to the increased effective number of degrees of freedom.

\begin{figure}
	\includegraphics[width=\columnwidth]{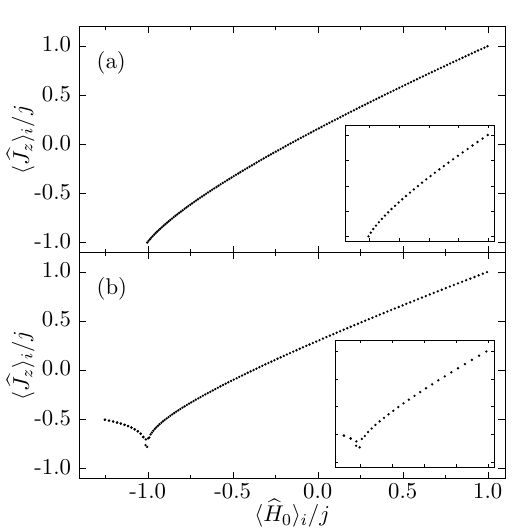}
	\caption{\label{fig:LH_PL}
    The Peres lattice for observables $\widehat{H}_0$ and $\widehat{J}_z$ for Hamiltonian \eqref{eq:LH} with (a) $\kappa = 0.7$ and (b) $\kappa = 2$.
    We set $j=70$ (the main panels) and $j = 15$ (the insets).}
\end{figure}

\begin{figure}
	\includegraphics[width=\columnwidth]{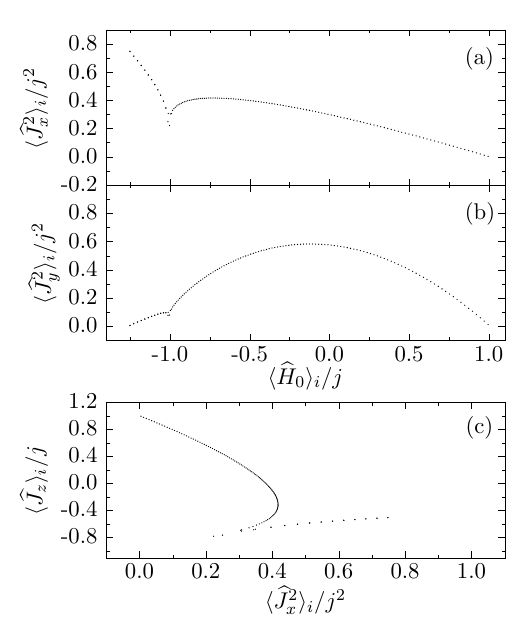}
	\caption{\label{fig:LHOO}
    Peres lattices for Hamiltonian \eqref{eq:LH} with some other choices of observables for $\kappa = 2$. We set $j = 70$.}
\end{figure}

Various Peres lattices of the stationary system are shown in Figs.\,\ref{fig:LH_PL} and \ref{fig:LHOO}.
While the lattices in Fig.\,\ref{fig:LH_PL} and those in Fig.\,\ref{fig:LHOO}(a--b) have on the abscissa the value $\langle\widehat{H}_0\rangle_i$ of the unperturbed energy $E_i$, the lattice in Fig.\,\ref{fig:LHOO}(c) shows correlations of some other quantities.
We note that the expectation values of sole observables $\widehat{J}_x$ and $\widehat{J}_y$ in the eigenstates $\ket{E_i}$ vanish due to the conservation of parity \eqref{eq:parit}, thus we can only use the squares (or other even powers) of these operators. 
In all cases, the coordinates of a given point in the Peres lattice hint at the location of the corresponding eigenstate on the Bloch sphere.
The excited-state quantum phase transition at $E=E_c$ creates the apparent cusp-shaped singularities in all lattices.
The cusp points vertically to the critical energy $E_c$ in the three lattices with the $\langle\widehat{H}_0\rangle_i$ abscissa, showing the convergence $\langle J_z\rangle_i/j \to -1$ and $\langle J^2_x\rangle_i/j^2,\langle J^2_y\rangle_i/j^2\to 0$ as $j\to\infty$.
In contrast, the cusp in the lattice of Fig.\,\ref{fig:LHOO}(c) is tilted, again pointing to the above-indicated limiting values of $\langle J^2_x\rangle_i$ and $\langle J_z\rangle_i$.

It should be noted that the points of the above $\kappa>1$ lattices that satisfy $\braket{\widehat{H}_0}_i < -j$ form nearly degenerate doublets. 
These are not distinguished in Figs.\,\ref{fig:LH_PL} and \ref{fig:LHOO} because both points of each doublet are almost identical. 
This systematic degeneracy is a consequence of the conservation of parity \eqref{eq:parit} and its spontaneous breaking in the $\kappa>\kappa_c$, $E<E_{c}$ domain for $j\to\infty$. 
If parity is used as the lattice observable on the vertical axis, each doublet splits to values $\langle\widehat{\Pi}\rangle_i=+1$ and $\langle\widehat{\Pi}\rangle_{i+1}=-1$.

\section{Kicked system\label{sec:kick}}

In this section, we describe various results related to the Delta driving from Eq.\,\eqref{eq:g_Delta}.
The $\delta$-driven Lipkin--Meshkov--Glick system, also called the kicked top model, is subject to extensive study in the literature \cite{Anand2024b, Zou2022, Olsacher2022, Wang2021, Sieberer2019, Russomanno2017, Chaudhury2009, Haake1987}.
Let us note that for this particular type of driving our analysis of Peres lattices is more detailed.
In Sect.\,\ref{sec:VD} we compare results on the kicked systems with those on the continuously driven systems. 

\subsection{Poincaré maps and Peres lattices}

\begin{figure}
	\includegraphics[width=\columnwidth]{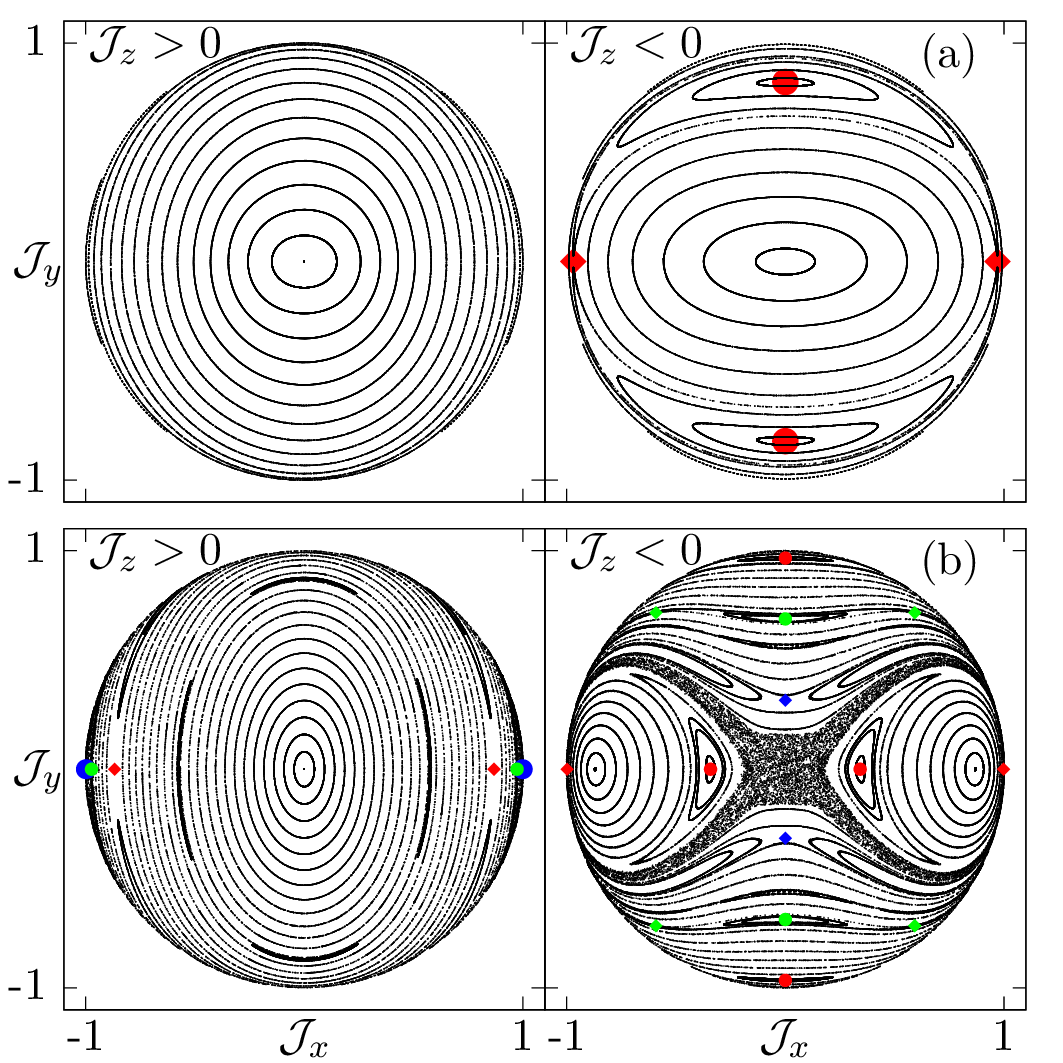}
	\caption{\label{fig:KT_CL_PR}
    Classical stroboscopic plot of the kicked system with $\mathcal{H}'=\mathcal{J}_z$. (a) The model with $\kappa = 0.7$, period $T = 7.15$, driving strength $\eta = 0.2$. (b) The model with $\kappa = 2$, period $T = 5$, driving strength $\eta = 0.01$. The colored points denote the elliptic (circles) and hyperbolic (diamonds) points associated with the correspondingly colored unperturbed trajectories in Fig.\,\ref{fig:LH_CL_PR}.}
\end{figure}

In Fig.\,\ref{fig:KT_CL_PR} we show classical stroboscopic maps of the kicked system with ${\kappa<\kappa_c}$ and ${\kappa>\kappa_c}$ for driving with ${\mathcal{H}'=\mathcal{J}_z}$ and some small values of driving strength~$\eta$. 
These maps, which are counterparts of the classical phase space images in Fig.\,\ref{fig:LH_CL_PR}, depict the locations of the classical kicked system on the Bloch sphere at times $t= nT$ for $n=0,1,2,3,\dots$.
They can be understood as classical Poincaré sections of the extended stationary system \eqref{eq:Hex} with the new coordinate fixed to $\phi\,{\rm mod}\,2\pi=0$ and the total conserved energy $\mathcal{H}_{\rm ext}$ set to a suitable constant.
We observe that the maps in Fig.\,\ref{fig:KT_CL_PR}(b) show some chaotic features, but a dominant part of the classical phase space remains regular for the small values of driving strength $\eta$ used here.
The most conspicuous features of these perturbed maps are families of periodic orbits and the associated resonance zones, which we discuss further in Sec.\,\ref{sec:reson}.

\begin{figure}
	\includegraphics[width=\columnwidth]{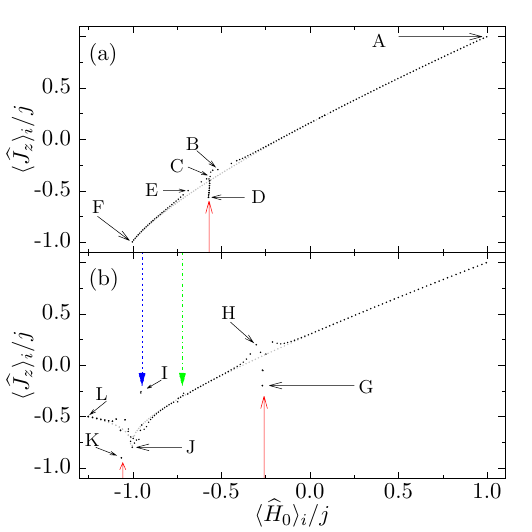}
	\caption{\label{fig:Double_plot_resonances} Peres lattices (black dots) for the kicked system with ${\widehat{H}'=\widehat{J}_z}$ and lattice observables $\widehat{H}_0\times\widehat{J}_z$. We set (a) ${\kappa = 0.7}$, ${\eta = 0.2}$, ${T = 7.15}$ and (b) ${\kappa = 2}$, ${\eta = 0.01}$, ${T = 5}$. In both cases ${j = 70}$. The background light-gray dots represent lattices of the stationary model (Fig.\,\ref{fig:LH_PL}). The colored vertical arrows correspond to the respective vertical lines in Fig. \ref{fig:Etau}. The letters link some Floquet modes to Figs. \ref{fig:OOI} and \ref{fig:HD_PL}.}
\end{figure}

\begin{figure}
	\includegraphics[width=\columnwidth]{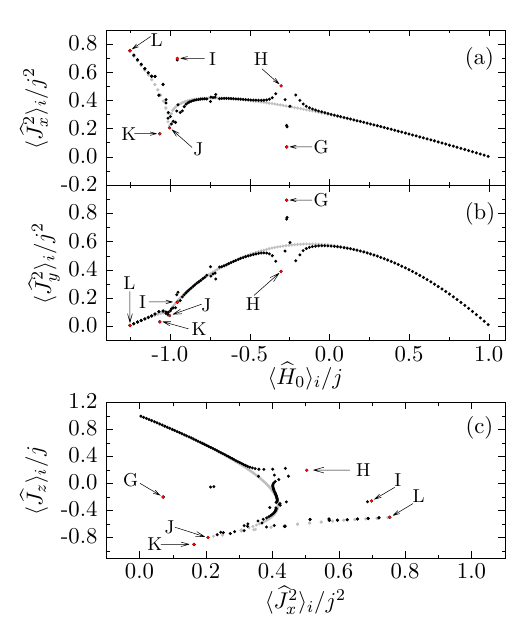}
	\caption{\label{fig:OOI}
    Peres lattices (black dots) of the $\widehat{J}_z$-kicked system for alternative lattice observables $\widehat{H}_0$, $\widehat{J}_z$, $\widehat{J}_x^2$ and $\widehat{J}_y^2$. We set ${\kappa = 2}$, ${\eta = 0.01}$, ${T = 5}$, and $j = 70$.  Light-grey dots correspond to the stationary model (Fig.\,\ref{fig:LHOO}); the letters denote the same modes as in Figs. \ref{fig:Double_plot_resonances} and \ref{fig:HD_PL}.}
\end{figure}

We compare the above Poincaré maps with their quantum counterparts\,---\,the corresponding Peres lattices of the kicked system.
As indicated above, these lattices must be constructed with the aid of the Floquet modes $\ket{F_i(t_0)}$, defined as eigenvectors of the Floquet operator $\widehat{U}(t_0+T, t_0)$ describing the quantum evolution of the driven system from a certain initial time $t_0$ over one period.
We have 
\begin{eqnarray}
	\widehat{U}(t_0+T, t_0)\ket{F_i(t_0)} = \mathrm{e}^{-\mathrm{i}F_iT} \ket{F_i(t_0)}, \label{eq:FloquetOperator}
\end{eqnarray}
where the eigenstates $\ket{F_i(t_0)}$ depend on $t_0$, while the complex eigenvalues, expressed via the so-called quasienergies $F_i \in \mathbb{R}$, do not.
The quasienergies are defined ambiguously, since adding to $F_i$ any integer multiple of $\frac{2\pi}{T}$ does not change the corresponding eigenvalue. 
This is solved by restricting the quasienergies to the first Brillouin zone defined as $F_i \in [-\frac{\pi}{T}, \frac{\pi}{T})$.
Therefore, the Peres lattice of a periodically driven system for lattice observables $\widehat{O}_1$ and $\widehat{O}_2$ is defined as a scatter plot of the expectation values $\bra{F_i(t_0)}\widehat{O}_1\ket{F_i(t_0)}\equiv\langle\widehat{O}_1\rangle_i$ versus $\bra{F_i(t_0)}\widehat{O}_2\ket{F_i(t_0)}\equiv\langle\widehat{O}_2\rangle_i$.
It is clear that the difference between the lattices for two starting times $t_0$ and $t'_0$ is expressed by the unitary transformation $\widehat{U}(t_0,t'_0)$ of both lattice operators $\widehat{O}_1$ and $\widehat{O}_2$.
Unless otherwise stated, we set ${t_0=0}$.
We further comment on the choice of initial time in Appendix \ref{sec:App_T_0}.

Peres lattices of the kicked system for the same lattice observables as in Figs.\,\ref{fig:LH_PL} and \ref{fig:LHOO} are presented in Figs.\,\ref{fig:Double_plot_resonances} and~\ref{fig:OOI}, respectively.
The parameters of the stationary Hamiltonian remain the same.
The driving operator is taken as $\widehat{H}'=\widehat{J}_z$, see Eq.\,\eqref{eq:Hprime_Jz}, and the driving strength $\eta$ and period $T$ are consistent with Fig.\,\ref{fig:KT_CL_PR}.
We note that numerical errors in all Peres lattices shown in this paper are under the visual resolution level. 
A note on the numerical precision can be found in Appendix \ref{sec:App_num}.

In view of the lattices in Figs.\,\ref{fig:Double_plot_resonances} and~\ref{fig:OOI}, we can immediately draw two main conclusions:
First, we observe that all Peres lattices of the kicked system with small driving strengths show a great degree of overall similarity to the corresponding lattice of the stationary system.
However, at some particular places the linear chains of points of the unperturbed lattice start decomposing into more disordered structures.
These seeds of chaos are connected, as discussed in Sect.\,\ref{sec:reson}, with resonances of the periodic driving with internal dynamics of the unperturbed system. 
In Sect.\,\ref{sec:chaos} we will see that with increasing driving strength $\eta$, chaotic behavior gradually spreads throughout the lattice, distorting it (as well as the classical Poincaré maps).
Second, the cusp-like structures of the unperturbed $\kappa>\kappa_c$ lattices, connected with the excited-state quantum phase transition at $E=E_c$, are partly preserved in the weakly perturbed system, which is in accord with the observations of Refs.\,\cite{Bandyopadhyay2015, Bastidas2014b, Bastidas2014a}.
This conclusion will be further elaborated in Sect.\,\ref{sec:quasi}, where we will investigate the phase-space images of individual Floquet modes.

\begin{figure*}
	\includegraphics[width=\textwidth]{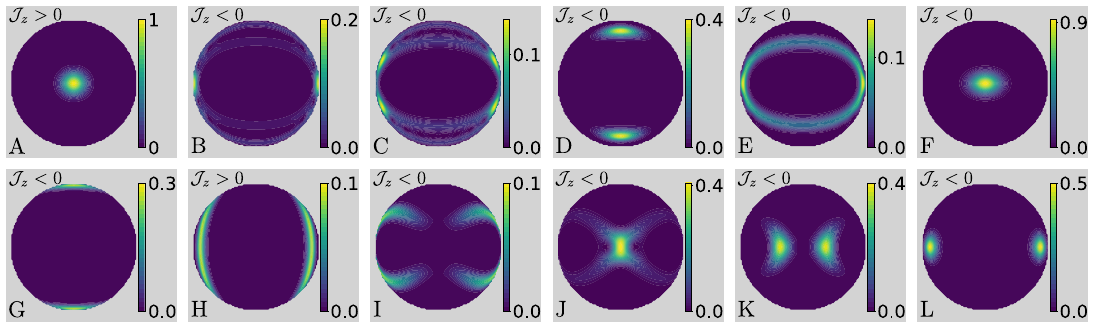}
	\caption{\label{fig:HD_PL}
    Husimi distributions \eqref{eq:Husi} corresponding to the Floquet states in Figs. \ref{fig:Double_plot_resonances} and \ref{fig:OOI} labeled with the respective letter A--L.  In each case, only one of the Bloch sphere hemispheres, the one with larger values of the Husimi distribution, is shown. The parameters are ${\kappa = 0.7}$, ${T = 7.15}$, ${\eta = 0.2}$ (upper panels) and ${\kappa = 2}$, ${T = 5}$, ${\eta = 0.01}$ (lower panels); $j = 70$.}
\end{figure*}

\subsection{Resonances}
\label{sec:reson}

The essential insight into the mechanism of the proliferation of chaos in the driven system is that it starts at the unperturbed energies $\mathcal{E}=\langle\widehat{H}_0\rangle_i/j$ satisfying the resonance condition
\begin{equation}\label{eq:res}
    \frac{\tau(\mathcal{E})}{T}=\frac{k}{l}
\end{equation}
with $k,l$ standing for small positive integers \cite{Reichl2004,Gutzwiller1990,Haake1987}.
This means that the external driving with period $T$ acts coherently with the period $\tau(\mathcal{E})$ of the internal dynamics. 
When we increase the driving strength~$\eta$ from zero, chains of elliptic and hyperbolic periodic points occur.
The hyperbolic periodic points represent the seeds of classical chaotic dynamics in the driven system \cite{Reichl2004}.
With further increase of~$\eta$, the resonance phase-space regions grow and start to overlap, creating domains of chaotic dynamics.
These features will be demonstrated in Sect.\,\ref{sec:chaos}. 
We note that the sensitivity of various resonant Floquet states with different values of $(k,l)$ from Eq.\,\eqref{eq:res} to the strength of the periodic perturbation was recently studied for a similar system in Ref.\,\cite{Landa2026a,Landa2026b}.

The resonance condition \eqref{eq:res} can be understood within the semiclassical theory of nonlinear resonances in weakly perturbed integrable systems \cite{Landa2026b, Wisniacki2015, Brodier2002, Mirbach1994}. 
The classical Hamiltonian in a vicinity of the resonance $(k,l)$ in a frame corotating with the resonant motion is approximated (after averaging over the time variable) by a pendulum-like Hamiltonian.
Its parameters depend on the driving type and strength, and can be inferred from the stroboscopic plot.
After quantization, this resonant Hamiltonian couples selected sequences of the unperturbed eigenstates into the relevant Floquet modes, whose structure can therefore be deduced.
This method in principle provides a quantitative description of the resonance effects that we demonstrate below in a purely phenomenological manner.

Some of the trajectories with periods $\tau(\mathcal{E})$ that yield resonant ratios with the period $T$ of the applied driving are highlighted in Fig.\,\ref{fig:LH_CL_PR}.
The phase-space domains in the vicinity of these trajectories are most affected by the driving, as seen in the Poincaré maps of the kicked system in Fig.\,\ref{fig:KT_CL_PR}, where we observe the emergence of some resonant periodic structures.
In particular, two elliptic and two hyperbolic periodic points (each having the return period~$T$) appear near the red trajectory in the connected region of the phase space for both choices of~$\kappa$. 
One elliptic and one hyperbolic periodic point (return period~$T$) occur near the red trajectories in each of the two confined regions of the phase space for~${\kappa=2}$.
Two elliptic and two hyperbolic periodic points (return period~$2T$) appear near the blue trajectory. 
Finally, four elliptic and four hyperbolic periodic points (return period~$4T$) emerge near the green trajectory.
All these periodic points are shown in color in Fig.\,\ref{fig:KT_CL_PR}.

The resonant ratios \eqref{eq:res} also imprint in the local distortions of the Peres lattices in Figs.\,\ref{fig:Double_plot_resonances} and~\ref{fig:OOI}.
The strongest resonances are marked by colored arrows in Fig.\,\ref{fig:Double_plot_resonances}, the corresponding periods being indicated in Fig.\,\ref{fig:Etau}.
According to Eq.\,\eqref{eq:res}, these resonances are associated with integer ratios ${(k,l)=(1,1)}$ (red), $(2,1)$ (blue) and $(4,3)$ (green).
It should be noted that for ${E<E_c}$, the unperturbed trajectories associated with the ratios $(2,1)$ and $(4,3)$ are located in the chaotic layer of the perturbed system; therefore, the corresponding resonances do not stand out in the lattice, and we do not mark them in Fig.\,\ref{fig:Double_plot_resonances}.
More resonance ratios $(k,l)$ become apparent when the Peres lattices become denser with an increasing value of $j$ (see Sect.\,\ref{sec:scalj}).
We point out that the strengths of individual resonances and the forms of the corresponding local distortions of the Peres lattice depend on the particular internal dynamics and the type of periodic driving (vulnerability of various orbits by specific external perturbations) \cite{Landa2026a, Haake1987}. 
Consequently, these features must be studied separately in each particular case.

\subsection{Husimi distributions}
\label{sec:quasi}

To obtain a phase-space image of individual Floquet modes, we use the Husimi distribution \cite{Pennini2013}
\begin{equation}\label{eq:Husi}
	Q(\vartheta, \varphi) = \left| \braket{\vartheta, \varphi| F_i} \right|^2,
\end{equation}
where $\ket{\vartheta,\varphi}$ is the coherent state defined by:
\begin{eqnarray}
	\ket{\vartheta,\varphi} = \frac{1}{\left(1 + |\zeta|^2\right)^j}\sum_{m = -j}^{+j} \left[\binom{2j}{m + j} \right]^\frac{1}{2} &\!\! \zeta^{j-m} \ket{m}, \qquad
    \\
    &\zeta  = \mathrm{e}^{\mathrm{i}\varphi}\tan\frac{\vartheta}{2},
    \nonumber
\end{eqnarray}
with $\ket{m}$ denoting the eigenstates of $\widehat{J}_z$ satisfying $\widehat{J}_z\ket{m}=m\ket{m}$.
The distribution \eqref{eq:Husi} displays the location of the Floquet state $\ket{F_i}$ on the Bloch sphere of Eq.\,\eqref{eq:Bloch}. 
This in general helps us to understand the nature and specific properties of this state.
For more details on the Husimi distributions, see Refs.\,\cite{Pennini2013,Honsa2022, Radcliffe1971, Husimi1940}.

To construct and inspect Husimi distributions for all Floquet states of the system would be too exhausting, but the method of Peres lattices gives us a tool for selecting the states that will be most interesting for various specific purposes we may have.  
In Fig. \ref{fig:HD_PL}, we show Husimi distributions of the Floquet states marked in Figs.\,\ref{fig:Double_plot_resonances} and~\ref{fig:OOI} by the corresponding labels A--L.

The forms of the Husimi distributions in Fig.\,\ref{fig:HD_PL} parallel the structures present in the classical Poincar{\'e} maps of the kicked system in Fig.\,\ref{fig:KT_CL_PR}. 
In particular, we clearly observe a greater localization of the Husimi distributions of the Floquet modes with various $\langle\widehat{H}_0\rangle_i$ at the periodic points of the Poincar{\'e} maps.
Floquet modes B, C and H, which are associated with the hyperbolic periodic points, and modes D, G, I, and K, associated with the elliptic periodic points, mark the distortions of the corresponding Peres lattices in Figs.\,\ref{fig:Double_plot_resonances} and~\ref{fig:OOI} from the unperturbed shapes.

The Floquet mode J in Fig.\,\ref{fig:HD_PL} is localized at the hyperbolic fixed point $(\mathcal{J}_x,\mathcal{J}_y,\mathcal{J}_z)=(0,0,-1)$ of the stationary system and shows an apparent similarity to the separatrix trajectory with $\mathcal{E}=-1$ in between the parity-breaking and parity-conserving phases, see Fig.\,\ref{fig:LH_CL_PR}.
In Fig.\,\ref{fig:Double_plot_resonances} we verify that this mode still forms the dip of the perturbed Peres lattice, similar to that in Fig.\,\ref{fig:LH_PL}.
We can therefore argue that the ${E=E_c}$ excited-state quantum phase transition of the stationary system with $\kappa>\kappa_c$ remains present in the ${\langle\widehat{H}_0\rangle_i=E_c}$ Floquet mode of the kicked system.
Similar conclusions are drawn in Refs.\,\cite{Bandyopadhyay2015, Bastidas2014b, Bastidas2014a}.
However, it needs to be stressed that this result holds only for not too large values of the driving strength $\eta$.

\subsection{Similarity to the unperturbed lattice}

Here, we quantify the overall similarity of Peres lattices of the kicked and stationary systems.
Considering the unperturbed lattice of observables $\widehat{H}_0\times\widehat{J}_z$, we introduce a continuous function $J_z(\mathcal{E})$ defined by
\begin{equation}\label{eq:inter}
    J_z(\mathcal{E})=\langle\widehat{J}_z\rangle_i
    \frac{E_{i+1}\!-\!j\mathcal{E}}{E_{i+1}\!-\!E_i}+
    \langle\widehat{J}_z\rangle_{i+1}
    \frac{j\mathcal{E}\!-\!E_i}{E_{i+1}\!-\!E_i} 
\end{equation}
for $j\mathcal{E}\in[E_i,E_{i+1}]$.
This function interpolates between individual points of the unperturbed lattice and becomes smoother as $j$ increases.
To avoid uncertainties associated with interpolation, it is convenient to evaluate $J_z(\mathcal{E})$ for a large enough value of $j$ (we do so for $j=10^4$).
In the limit $j\to\infty$, the value $J_z(\mathcal{E})$ coincides with the asymptotic-time average $\langle\mathcal{J}_z\rangle$ of classical quantity $\mathcal{J}_z$ over the trajectory with energy $\mathcal{E}$.
The similarity of the unperturbed lattice with the corresponding $\widehat{H}_0\times\widehat{J}_z$ lattice of the periodically driven system can then be quantified by a measure defined as follows: 
\begin{equation}\label{eq:measure}
	\mathcal{N} \!=\! \frac{1}{2j\!+\!1}\! \sum_{i = 1}^{2j + 1}\!\left[\!\frac{\bra{F_i}\widehat{J}_z\ket{F_i}}{j} \!-\! J_z\biggl(\!\mathcal{E}\!=\!\frac{\bra{F_i}\widehat{H}_0\ket{F_i}}{j}\biggr)\! \right]^2
    \!\!.
\end{equation}
This quantity is analogous to the mean squared error, known from statistics, thus a~low value of $\mathcal{N}$ indicates a~high degree of similarity between the perturbed and unperturbed lattices, and vice versa.
Let us point out that the measure \eqref{eq:measure} does not take into account deviations of the lattice points in the horizontal direction, which, however, may play a non-negligible role only for small values of $j$. 

\begin{figure}
	\includegraphics[width=\columnwidth]{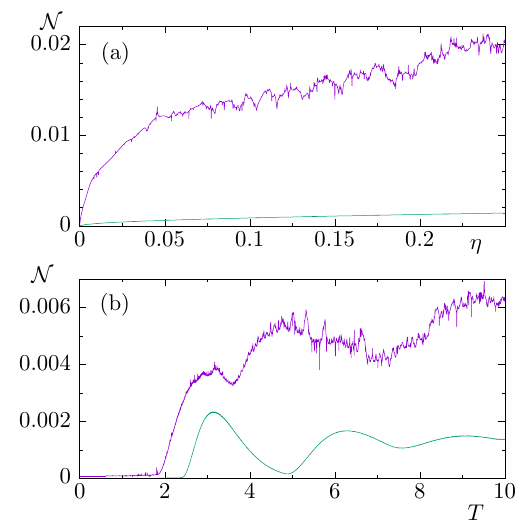}
	\caption{\label{fig:MSE} The dependence of the dissimilarity measure \eqref{eq:measure} on the driving parameters for the $\widehat{J}_z$-kicked system. (a) Dependence on the driving strength $\eta$ for ${\kappa = 0.7}$, ${T = 7.15}$ (green line) and ${\kappa = 2}$, ${T = 5}$ (purple line). (b) Dependence on the driving period $T$ for ${\kappa = 0.7}$, ${\eta = 0.2}$ (green line) and ${\kappa = 2}$, ${\eta = 0.01}$ (purple line). In both plots, we set $j=80$.}
\end{figure}

The dependencies of the measure $\mathcal{N}$ on the driving strength $\eta$ and period $T$ are depicted in Fig. \ref{fig:MSE}.
We observe a non-monotonously increasing value of $\mathcal{N}$ with both increasing $\eta$ and $T$, i.e., a non-monotonously decreasing similarity of the perturbed and unperturbed lattices.
The dependence on the period $T$ shows a threshold under which $\mathcal{N}\approx 0$. 
This results from the absence of strong resonances in the perturbed lattices with ${T\lesssim 2}$. 
A glance at Fig.\,\ref{fig:Etau} confirms this observation, the shortest classical period for ${\kappa = 0.7}$ is ${\tau \approx 4.8}$ and for ${\kappa = 2}$ it is ${\tau \approx 3.6}$. Consequently, the (2, 1) resonance first occurs for ${T \approx 2.4}$ and ${T \approx 1.8}$ for our two sets of parameters, respectively.
We also see that $\mathcal{N}$ is significantly smaller in both dependencies for ${\kappa<\kappa_c}$ than for ${\kappa>\kappa_c}$.
This is a consequence of more efficient resonance distortions in the ${\kappa>\kappa_c}$ lattice.
Indeed, the hyperbolic fixed point ${\mathcal{E}=-1}$ with ${\tau\to\infty}$ for ${\kappa>\kappa_c}$, see Fig.\,\ref{fig:Etau}, induces pairwise arranged resonance values of $\langle\widehat{H}_0\rangle_i$, which enhances the vulnerability of the system under the periodic perturbation.

\begin{figure*}
	\includegraphics{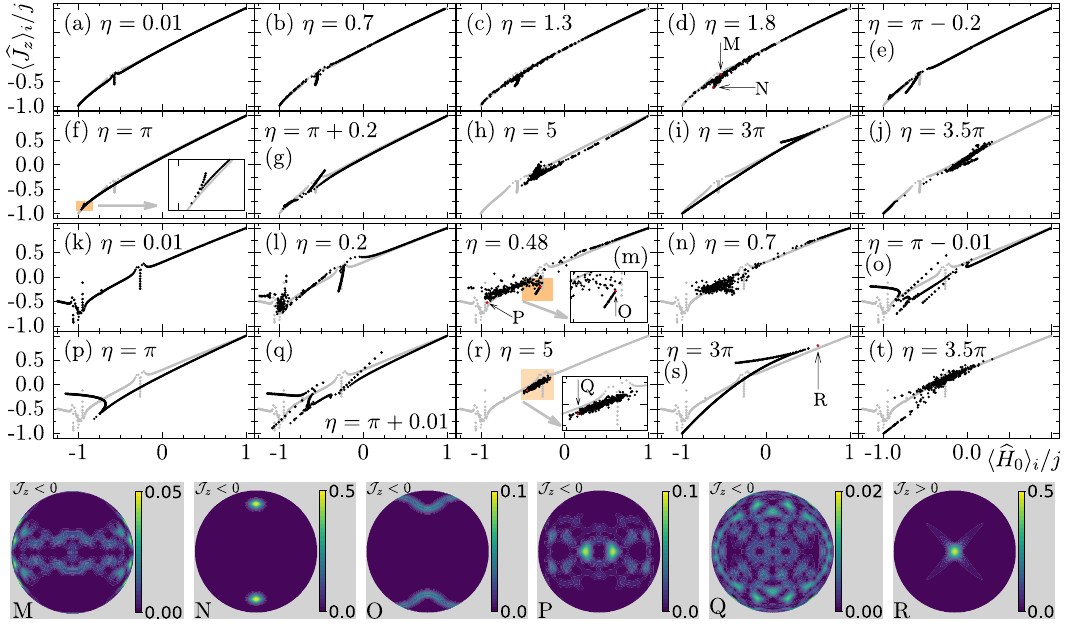}
	\caption{\label{fig:EV}
    The $\widehat{H}_0\times\widehat{J}_z$ Peres lattices of the $\widehat{J}_z$-kicked system for various values of the driving strength $\eta$. The parameters are ${\kappa = 0.7, T = 7.15}$ in panels (a)--(j) and ${\kappa = 2, T = 5}$ in panels (k)--(t). The value of $\eta$ is indicated in each plot. For comparison, the lattice for ${\eta = 0.2}$ in panels (a)--(j) and ${\eta = 0.01}$ in panels (k)--(t) is shown in gray in the background. In all panels we set ${j = 200}$. Panels M--R show Husimi distributions \eqref{eq:Husi} of selected Floquet modes, whose lattice points (red) are indicated by arrows in panels (a)--(t). For all these modes except Q, the population of the hidden Bloch hemisphere is negligible.}
\end{figure*}

\subsection{Regular and chaotic regimes}
\label{sec:chaos}

So far, we have discussed a weak perturbation regime with small values of the driving strength $\eta$. 
Now we examine the regime of larger $\eta$, in which we anticipate dynamics with a higher degree of chaos.
On the quantum level, this anticipation follows from the evolution operator of the kicked system over a single period,
\begin{eqnarray}
	\widehat{U}(T, 0) = e^{-\frac{\mathrm{i}}{2}(\widehat{H}_0 T - \eta \widehat{J}_z)} e^{-\mathrm{i} \eta \widehat{J}_z} e^{-\frac{\mathrm{i}}{2}(\widehat{H}_0 T - \eta \widehat{J}_z)}, \label{eq:UKT}
\end{eqnarray}
which with increasing $\eta$ more and more deviates from the unperturbed (regular) form ${\widehat{U}_0(T, 0) = e^{-\mathrm{i}\widehat{H}_0 T}}$.

However, this general anticipation is verified only within some finite intervals of $\eta$. 
A more careful analysis reveals that the fully regular dynamics reappears periodically even for very large values of $\eta$.
This follows from the observation that the transformation 
\begin{equation}
    (\eta, \kappa, T) \mapsto (\eta', \kappa', T') 
    \!=\! \left(\eta\!+\!n\pi,\frac{\kappa T}{T + n \pi},T\!+\!n \pi \right)
    \label{eq:parasym}
\end{equation}
with any $n \in \mathbb{Z}$ leaves the evolution operator (\ref{eq:UKT}) unchanged up to a phase.
Indeed, expressing the stationary Hamiltonian $\widehat{H}_0$ from Eq.\,\eqref{eq:LH}, we find that the evolution operator for parameters $(\eta', \kappa', T')$ reduces to
\begin{equation}
	\widehat{U}'(T', 0) = \left\{\begin{array}{ll}
    \widehat{U}(T, 0) & \mathrm{for}\ n\ \mathrm{even,}\\
    \widehat{U}(T, 0)\hat{\Pi}(-1)^{-j} & \mathrm{for}\ n\ \mathrm{odd,}
    \end{array}\right.
    \label{eq:UKTprime}
\end{equation}
where the conserved parity operator \eqref{eq:parit} and the $(-1)^{-j}$ term generate just a phase factor in the odd-$n$ case if the initial state has a fixed parity.
Thus, when the increasing driving strength passes through the points ${\eta=\pi,2\pi,3\pi\,\dots}$, the system with the given values of parameters $\kappa$ and $T$ becomes equivalent to the fully regular ${\eta=0}$ system with a sequence of modified interaction parameters ${\kappa'}$ from Eq.\,\eqref{eq:parasym} with ${n=-1,-2,-3,\dots}$

In panels (a)--(t) of Fig.\,\ref{fig:EV}, we present examples of Peres lattices with the two choices of parameter $\kappa$ for several values of the driving strength $\eta$. 
We see that the disturbances of the regular ${\eta=0}$ lattice first occur at the resonance points and grow with increasing~$\eta$.
In some cases, they merge into whole regions of disordered points corresponding to completely chaotic dynamics.

Chaos emerges in the regions where the Husimi distributions of the concerned Floquet modes overlap with the hyperbolic periodic points of the classical Poincar{\'e} map.
This is demonstrated by the Husimi distributions of selected Floquet modes shown in panels M--R of Fig.\,\ref{fig:EV}, in which increasing chaoticity shows up in spreading of the Husimi distribution over larger domains of the phase space.
Panels N and M, respectively, exemplify regular and chaotic modes for ${\kappa<\kappa_c}$, while panels O and R, and P and Q, respectively, represent regular and chaotic modes for ${\kappa>\kappa_c}$.
We verify that the points corresponding to the modes with regular or chaotic Husimi distribution in the relevant Peres lattice\,---\,see the arrows in panels (c), (m), (r) and (s)\,---\,belong, respectively, to the ordered or disordered fractions of the lattice. 

When the degree of chaos becomes high, the points in the Peres lattices begin to shrink, as exemplified in Fig.\,\ref{fig:EV}(r). 
This is a sign of ergodicity. 
Indeed, classical chaotic trajectories cover the entire available domain of the phase space (Husimi distributions of the corresponding Floquet modes spread out over this domain, see for instance the mode Q in Fig.\,\ref{fig:EV}), thus the time averages of observables for individual trajectories (or quantum expectation values for individual Floquet modes) can be well approximated by the uniform phase-space averages \cite{Russomanno2015, Hortikar1998, Berry1977b}.

Figure~\ref{fig:EV} also demonstrates resurgences of full regularity at the above-discussed values ${\eta=n\pi}$.
Indeed, near these values, all signatures of chaos are suppressed and the lattice again takes regular forms, its particular arrangement being varied with the individual integer values of $n$. 
The cases ${n=1}$ and ${n=3}$ are shown in panels (f) and (p), and (i) and (s), respectively. 
The Husimi distribution of the regular Floquet mode R on the tip of lattice (s) indicates the formation of a new hyperbolic periodic point on the upper Bloch hemisphere.

\subsection{Increasing the system size}
\label{sec:scalj}

How do the Peres lattices change when the size parameter $j$ increases?
Since the number of lattice points is equal to the Hilbert space dimension ${2j+1}$, the lattices become denser and reveal finer details and substructures.
The high-$j$ lattices tend to express correlations between classical time averages of relevant observables over individual trajectories.
However, due to nontrivial size-dependent scaling of the sensitivity of individual Floquet modes to perturbation \cite{Landa2026a}, the Peres lattices with increasing $j$ may change some of their qualitative features.

\begin{figure}
	\includegraphics[width=\columnwidth]{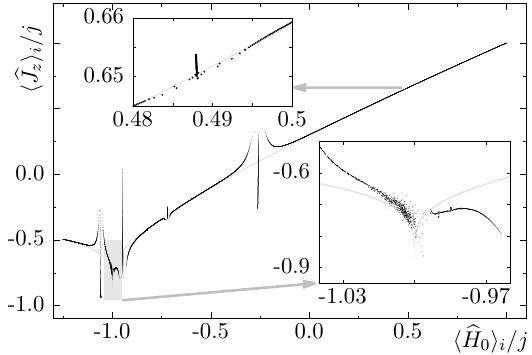}
	\caption{\label{fig:big}Normalized $\widehat{H}_0\times\widehat{J}_z$ Peres lattice of the $\widehat{J}_z$-kicked system for ${j = 10\,000}$ and ${\kappa = 2,\eta = 0.01,T = 5}$. The insets enlarge the selected parts of the lattice. The gray line in the background represents the corresponding lattice of the stationary system.}
\end{figure}

As an example, the $\widehat{H}_0\times\widehat{J}_z$ Peres lattice for ${j = 10^4}$ is depicted in Fig.\,\ref{fig:big}.
It is clear that the lattice points corresponding to regular trajectories are now so dense that they seem to form almost a continuous shape. 
However, zooming in, we distinguish small resonances, which were not apparent in the lower panel of Fig.\,\ref{fig:Double_plot_resonances} showing the same lattice for a lower $j$, for instance the resonance around ${\braket{\widehat{H}_0}_i/j = 0.49}$ (upper inset of Fig.\,\ref{fig:big}).
Also the fully disordered parts of the lattice corresponding to strongly chaotic dynamics are more clearly defined (see the lower inset of Fig.\,\ref{fig:big}). 
We observe that such chaotic parts are more compact for high $j$, the oscillations around the unperturbed form of the lattice being reduced in comparison to low-$j$ cases.
This is again a~consequence of ergodicity (cf.\,Sec.\,\ref{sec:chaos}), which tends to contract the extent of the chaotic domains in the Peres lattice.
In the regime of small driving strength $\eta$, the contraction is apparently stronger along the $\langle\widehat{J}_z\rangle_i$ axis than along $\langle\widehat{H}_0\rangle_i$.

\subsection{Changing the driving operator}

\begin{figure}
	\includegraphics[width=\columnwidth]{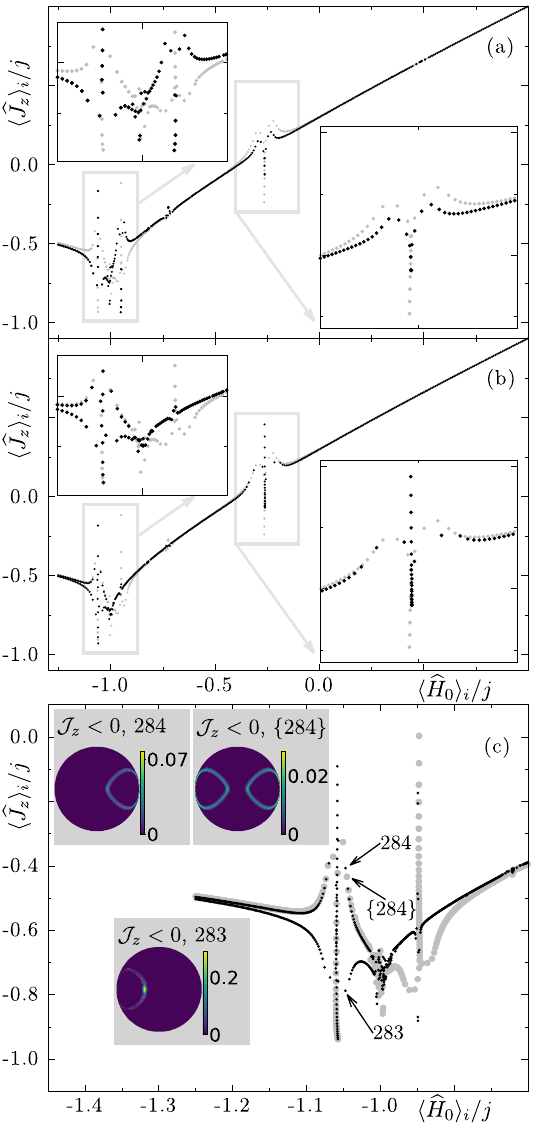}
	\caption{\label{fig:ODO}The $\widehat{H}_0\times\widehat{J}_z$ Peres lattice of the kicked system for various driving operators, namely ${\widehat{H}^\prime=\widehat{J}_z^2/j}$ in panel (a), and ${\widehat{H}^\prime=\widehat{J}_x}$ in panels (b) and (c). For comparison, the lattice for the driving operator $\widehat{J}_z$ is shown in gray in the background. The parameters are ${\kappa = 2, T= 5, \eta = 0.01}$ for all panels, ${j = 200}$ for panels (a) and (b), and ${j = 1000}$ for panel (c). The insets in panel (c) show the Husimi distributions of selected Floquet modes, namely the 283rd mode in the lower inset and the 284th mode in the upper-left inset (ordering with respect to the expectation value $\langle\widehat{H}_0\rangle_i$), both corresponding to ${\widehat{H}^\prime = \widehat{J}_x}$. For comparison, the upper-right inset shows the 284th mode for ${\widehat{H}^\prime = \widehat{J}_z}$.}
\end{figure}

So far we have studied the $\delta$-kicked system with the driving operator ${\widehat{H}^\prime = \widehat{J}_z}$ from Eq.\,\eqref{eq:Hprime_Jz}.
Here we examine how the Peres lattice of the kicked system changes when the driving operator is altered.
In particular, we consider the driving operators ${\widehat{H}^\prime = \widehat{J}_z^2/j}$ and ${\widehat{H}^\prime = \widehat{J}_x}$ from Eqs.\,\eqref{eq:Hprime_squareJz} and \eqref{eq:Hprime_Jx}. 

Peres lattices corresponding to both these new choices of the driving operator are shown in Fig.\,\ref{fig:ODO}. 
We may immediately say that the overall shape of the lattice remains the same as in the ${\widehat{H}^\prime = \widehat{J}_z}$ case, but the patterns in the resonance regions are different.
This is not surprising, as we are dealing here again with only a weakly perturbed system, whose Peres lattices differ from the unperturbed ones in the resonance regions only.

However, there is an essential difference between the choices ${\widehat{H}^\prime = \widehat{J}_z}$ and ${\widehat{H}^\prime = \widehat{J}_z^2/j}$, on the one hand, and ${\widehat{H}^\prime = \widehat{J}_x}$,
on the other:
While for the former choices the parity~\eqref{eq:parit} is conserved, for the latter choice it is not.
As a consequence, the points with ${\braket{\widehat{H}_0}_i < -j}$ in the ${\kappa>\kappa_c}$ lattices corresponding to ${\widehat{H}^\prime = \widehat{J}_z}$ and ${\widehat{H}^\prime = \widehat{J}_z^2/j}$ are actually parity doublets\,---\,pairs of points with opposite parities that lie very close to each other.  
The Floquet modes corresponding to both positive- and negative-parity members of each doublet have Husimi distributions localized simultaneously in both confined phase-space regions of the stationary system (Fig.\,\ref{fig:LH_CL_PR}).  

In contrast, the doublets are absent in the Peres lattice for the parity-violating driving operator ${\widehat{H}^\prime = \widehat{J}_x}$. 
In Fig.\,\ref{fig:ODO}(c) we see that what was previously (in the parity-conserving cases) a single string of points with a resonance represented by a more or less straight vertical line segment splits now (in the parity-violating case) into two strings of points and two superimposed line segments.
We have verified (see below) that the upper and lower strings of points correspond, respectively, to the Floquet modes localized in the right- and left-side confined phase-space lobes of the unperturbed system (Fig.\,\ref{fig:LH_CL_PR}). 
The strikingly different forms of the Peres lattices manifest important features of the parity-conserving and parity-violating regimes of driving.

The insets of Fig\ \ref{fig:ODO}(c) show the Husimi distribution of the 283rd and 284th Floquet modes from the lower and upper strings of the system with ${\widehat{H}^\prime = \widehat{J}_x}$. 
For comparison, we also show the Husimi distribution for the 284th Floquet mode (denoted by the symbol $\{284\}$) of the system with ${\widehat{H}^\prime = \widehat{J}_z}$.
We note that the Husimi distribution located in the right-side phase-space lobe for the $\widehat{J}_x$-kicked system is very similar to the right-side part of the symmetric Husimi distribution for the $\widehat{J}_z$-kicked system.
This is why the upper branch of points in the Peres lattice of the $\widehat{J}_x$-kicked system in Fig.\,\ref{fig:ODO}(c) is almost identical to the lattice of the $\widehat{J}_z$-kicked system.
In contrast, Husimi distributions in the left-side phase-space lobe of the $\widehat{J}_x$-kicked system differ from their counterparts in the $\widehat{J}_z$-kicked system and consequently the corresponding points in the Peres lattice are displaced.

\subsection{Probing the Floquet Hamiltonian
\label{sec:BCH}}

To describe dynamics of general periodically driven systems, it is useful to introduce a so-called Floquet Hamiltonian $\widehat{G}$. 
It is defined by the following expression of the evolution operator,
\begin{eqnarray}
	\widehat{U}(t_1, t_0) = 
    \mathrm{e}^{-\mathrm{i}\widehat{K} (t_1)} 
    \mathrm{e}^{-\mathrm{i}(t_1 - t_0) \widehat{G}} 
    \mathrm{e}^{\mathrm{i}\widehat{K}(t_0)}, \label{eq:FloquetHamiltonian}
\end{eqnarray}
where $\widehat{U}(t_1, t_0)$ evolves the system between general times $t_0$ and~$t_1$, and $\widehat{K}(t)$ is a Hermitian time-dependent operator (so-called kick operator) satisfying ${\widehat{K}(t + T) = \widehat{K}(t)}$ \cite{Bukov2015,Goldman2014,Shirley1965}. 
The time-independent Hermitian operator $\widehat{G}$ plays the role of a stationary Hamiltonian that generates the evolution of the system, up to the unitary transformations expressed by $\widehat{K}(t_0)$ and $\widehat{K}(t_1)$. 

In the setting used here (see Sect.\,\ref{sec:ham}), the Floquet Hamiltonian $\widehat{G}$ and the corresponding kick operator $\widehat{K}(t)$ can be obtained as series in powers of parameters $\eta$ and $T$, namely
\begin{equation}
\begin{array}{c}
    \widehat{G}=\lim\limits_{N\to\infty}\widehat{G}^{(N)},\qquad
    \widehat{K}(t)=\lim\limits_{N\to\infty}\widehat{K}(t)^{(N)},
    \\ 
    \widehat{G}^{(N)}=\sum\limits_{n=0}^N
    \!\!\!\sum\limits_{\begin{smallmatrix}n_1,n_2\\n_1+n_2=n\end{smallmatrix}}\!\!\!\!\!\widehat{G}_{n_1n_2}\, \eta^{n_1}T^{n_2},
    \\
    \widehat{K}(t)^{(N)}=\sum\limits_{n=1}^N
    \!\!\!\sum\limits_{\begin{smallmatrix}n_1,n_2\\n_1+n_2=n\end{smallmatrix}}\!\!\!\!\!\widehat{K}(t)_{n_1n_2}\, \eta^{n_1}T^{n_2},
    \end{array}
    \label{eq:expansions}
\end{equation}
where $\widehat{G}_{n_1n_2}$ and $\widehat{K}(t)_{n_1n_2}$ are some operators that need to be determined for each particular type of periodic driving (see also Sect.\,\ref{sec:EH} below).
The series generally converges for low enough values of parameters $\eta$ and $T$, and in such cases it is natural to assume that the truncated operators $\widehat{G}^{(N)}$ and $\widehat{K}(t)^{(N)}$ with a sufficiently high $N$ represent a good approximation of the full Floquet operator $\widehat{G}$ and the corresponding $\widehat{K}(t)$.
However, for the present choice of parameters, convergence is neither guaranteed nor generally expected \cite{Haga2019, Biagi2018, Kuwahara2016, Blanes2009, Casas2009, Casas2007, Blanes2004}, and our results indeed strongly indicate its absence.

In the case of the $\delta$-kicked system, the method for constructing the Floquet Hamiltonian and the associated unitary transformations is rather straightforward.
It relies on the application of the Baker–Campbell–Hausdorff (BCH) expansion.
In particular, assuming the initial time just before the kick, i.e., ${t_0=\frac{T}{2}-\epsilon}$, where ${0<\epsilon\ll T}$, and the final time after one period just before the next kick, ${t_1 = t_0 + T=\frac{3T}{2}-\epsilon}$, the determination of $\widehat{G}$ reduces to finding the logarithm of product of exponentials on the right-hand side of  the expression
\begin{equation}
	\mathrm{e}^{-\mathrm{i}\widehat{G}T} = 
    \mathrm{e}^{-\mathrm{i} \left(\widehat{H}_0 - \frac{\eta}{T} \widehat{H}' \right) T} 
    \mathrm{e}^{-\mathrm{i} \eta \widehat{H}'}.
    \label{eq:BCHdef}
\end{equation}
For these particular initial and final times we have ${\widehat{K}(t_0)=\widehat{K}(t_1)=0}$, thus ${\widehat{U}(t_1, t_0) = \mathrm{e}^{-\mathrm{i}\widehat{G}T}}$.
To determine the evolution operator \eqref{eq:FloquetHamiltonian} between general times, the initial- and final-time unitary transformations can be obtained from a formally similar expression
\begin{equation}
	\mathrm{e}^{\mathrm{i}\widehat{K}(t)} = 
    \mathrm{e}^{\mathrm{i} \widehat{G} \left(\tfrac{T}{2}-t\right)\mathrm{mod}T} 
    \mathrm{e}^{-\mathrm{i} \left(\widehat{H}_0 - \frac{\eta}{T} \widehat{H}' \right) \left(\tfrac{T}{2}-t\right)\mathrm{mod}T},
    \label{eq:BCHdef2}
\end{equation}
where the function $\widehat{K}(t)$ again follows from the logarithm of the product of exponentials on the right-hand side. 
We note that the unitary transformation of the Floquet modes between the present (${t_0=\frac{T}{2}-\epsilon}$) and previous (${t_0=0}$) choices of the initial time leads to a slight modification of the corresponding Husimi distributions (as well as the classical return maps).
The corresponding Peres lattices are very similar, the resonances being located at the same places, although possibly taking slightly different shapes.

The BCH formula provides an expansion for the operator $\widehat{C}$ in the formula $\mathrm{e}^{\widehat{A}} \mathrm{e}^{\widehat{B}}=\mathrm{e}^{\widehat{C}}$ in terms of recursive commutators of operators $\widehat{A}$ and $\widehat{B}$: 
\begin{equation}
	\widehat{C}\!=\!
    \widehat{A}+\widehat{B}+\frac{1}{2}[\widehat A, \widehat B]+\frac{1}{12} [\widehat{A}, [\widehat  A, \widehat B]]+\frac{1}{12} [\widehat{B},[\widehat  B, \widehat A]]+\dots
	\label{eq:BCH_C}
\end{equation}
This series converges if the operators $\widehat{A}$ and $\widehat{B}$ are sufficiently small. 
Various convergence criteria have been formulated in the literature \cite{Haga2019, Biagi2018, Kuwahara2016, Blanes2009, Casas2009, Casas2007, Blanes2004}, most of them providing only a sufficient and not necessary condition.

\begin{figure}
	\includegraphics[width=\columnwidth]{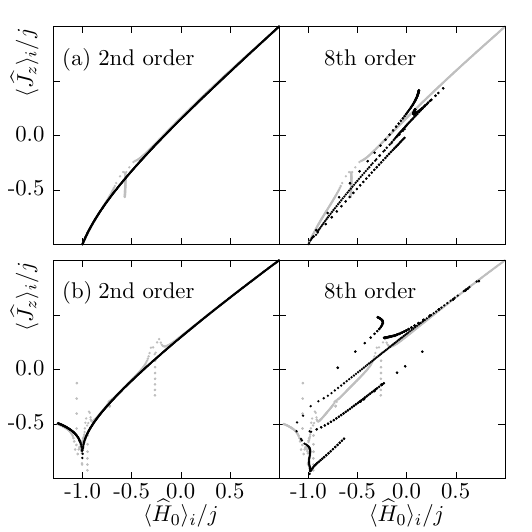}
	\caption{\label{fig:BCH}
    The $\widehat{H}_0\times\widehat{J}_z$ Peres lattices for the eigenvectors of the approximate Floquet Hamiltonian constructed by the BCH expansion of Eq.\,\eqref{eq:BCHdef} truncated at the second (left) and eighth  (right) orders for the $\widehat{J}_z$-kicked system. The parameters are (a) $\kappa = 0.7, \eta = 0.2, T = 7.15, j=200$ (top) and (b) $\kappa = 2, \eta = 0.01, T = 5, j=200$ (bottom). The corresponding Peres lattices of exact Floquet modes for the initial time ${t_0=T/2 - \epsilon}$ are shown by gray dots in the background. The lattices were computed in octuple precision to avoid numerical errors.
}
\end{figure}

The approximation of the Floquet Hamiltonian $\widehat{G}$ from Eq.\,\eqref{eq:BCHdef} for ${\widehat{H}^\prime = \widehat{J}_z}$ obtained by truncating the BCH expansion \eqref{eq:BCH_C} at the second order (including terms with up to two nested commutators) reads as follows:
\begin{eqnarray}
	\widehat{G}^{(2)} &=& \widehat{J}_z - \frac{\kappa}{2 j} \widehat{J}_x^2 + \frac{\kappa \eta}{4 j} \left(\widehat{J}_x \widehat{J}_y + \widehat{J}_y \widehat{J}_x \right) 
    \nonumber\\
	&-& \frac{\kappa T \eta}{12 j}\left(1-\frac{2\eta}{T}\right) \left( \widehat{J}_x^2 - \widehat{J}_y^2 \right) \nonumber\\
	&-& \frac{\kappa^2 T \eta}{48 j^2} \left( \widehat{J}_x^2 \widehat{J}_z + 2 \widehat{J}_x \widehat{J}_z \widehat{J}_x + \widehat{J}_z \widehat{J}_x^2 \right).
	\label{eq:G2BCH}
\end{eqnarray}
The structure of this expression corresponds to Eq.\,\eqref{eq:expansions}.
The corresponding truncated operator function $\widehat{K}(t)^{(2)}$ can, in principle, be determined by applying the same BCH expansion to Eq.\,\eqref{eq:BCHdef2}.
We have calculated the expansion of the Floquet Hamiltonian to much higher orders (up to ${N=18}$), with the resulting expressions for $\widehat{G}^{(N)}$ not explicitly given here.
The eigenvectors of $\widehat{G}^{(N)}$, which coincide with the truncated Floquet modes $\ket{F_i^{(N)}}$ corresponding to the evolution operator $\widehat{U}^{(N)}(\frac{3T}{2}-\epsilon,\frac{T}{2}-\epsilon) = \mathrm{e}^{-\mathrm{i} \widehat{G}^{(N)}T}$, can be used to construct the Peres lattices.
We compare the $\widehat{H}_0\times\widehat{J}_z$ Peres lattices for the ${N=2}$ and ${N=8}$ truncated modes with the lattice for the exact Floquet modes in Fig. \ref{fig:BCH}.

We see that whereas the second-order expansion captures the overall shape of the exact lattices, without the resonances, the eighth-order expansion completely fails to mimic even the overall shape.
Our numerical analysis reveals that the lattices start to depart significantly from the unperturbed shape already at ${N=6}$.
This suggests that the BCH expansion is strongly divergent for the present sets of parameters.
Indeed, we checked that Peres lattices up to the ${N=18}$ approximation (constructed for a smaller size ${j = 30}$) still exhibit significant and persistent deviations from the exact lattice.
As shown in Appendix \ref{sec:App_Convergence}, the convergence of the BCH expansion can be resumed by decreasing the driving period $T$ or the system size $j$, but in these convergent cases the resonances virtually disappear.

\section{Continuously driven system\label{sec:VD}}

In this section, we present results related to continuously driven systems with periodic driving functions from Eqs.\,\eqref{eq:g_Cos1}--\eqref{eq:g_Tent}, see Fig.\,\ref{fig:gs}.
Note that the Delta driving (\ref{eq:g_Delta}) can be represented by the Fourier series
\begin{eqnarray}
	-\frac{1}{T}+\delta(t-T/2) = \frac{2}{T} \sum_{n = 1}^{+\infty} (-1)^n \cos{\frac{2 \pi n t}{T}},
    \label{eq:expdelta}
\end{eqnarray}
hence, the Cos1 and Cos2 drivings (\ref{eq:g_Cos1}) and (\ref{eq:g_Cos2}) represent, respectively, the first one and two terms of this series.
Also the other types of continuous drivings are arranged so that they can be related to the Delta driving, see Fig.\,\ref{fig:gs}.
Since, for weak or medium strengths $\eta$, the results for continuously driven systems are rather similar to those for the kicked system, our present discussion will be less detailed than that in Sect.\,\ref{sec:kick}.

\subsection{Peres lattices\label{sec:PL_VD}}

\begin{figure}
	\includegraphics{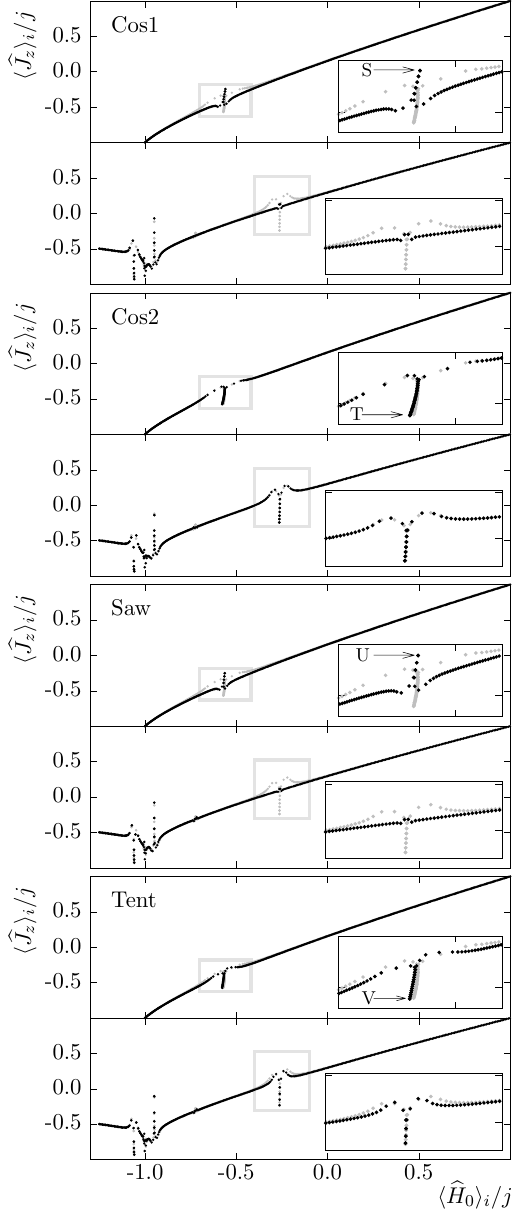}
	\caption{\label{fig:VD}
    The $\widehat{H}_0\times\widehat{J}_z$ Peres lattices for the continuous driving functions Cos1, Cos2, Saw and Tent in Fig.\,\ref{fig:gs}. Each driving form is represented by two lattices corresponding to two sets of parameters: ${\kappa = 0.7, T = 7.15, \eta = 0.2, j = 200}$ (upper panels) and ${\kappa = 2, T = 5, \eta = 0.01, j = 200}$ (lower panels). The gray background dots depict the respective lattices for the Delta driving. The Floquet modes S, T, U and V marked by arrows are visualized in Fig.\,\ref{fig:VD_Husimi}.}
\end{figure}

The $\widehat{H}_0\times\widehat{J}_z$ Peres lattices for the four types of continuous driving and for both parameter sets used in Sect.\,\ref{sec:kick} are depicted in Fig.\,\ref{fig:VD}.
The lattices corresponding to the Delta driving with the same parameters are shown by gray dots in the background for comparison.
We see that all the presently employed continuous driving forms preserve the positions of the main resonances with respect to the Delta driving.
The Cos2 and Tent driving functions in Fig.\,\ref{fig:gs} are visually closer to the Delta driving than the Cos1 and Saw driving functions, which is also reflected by the corresponding Peres lattices:
For the Cos2 and Tent drivings, the shapes of the lattices remain practically the same as in the Delta case, while for the Cos1 and Saw drivings, the shapes near the resonances are distorted. 
This conclusion is supported by similar results obtained for other choices of the lattice and driving operators.

\begin{figure}
	\includegraphics{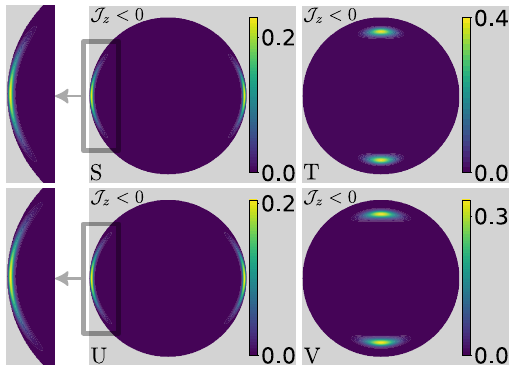}
	\caption{\label{fig:VD_Husimi}
    The Husimi distributions \eqref{eq:Husi} of the selected Floquet modes S, T, U and V from Fig.\,\ref{fig:VD}. Only the lower Bloch hemisphere is shown, the values of $Q$ on the other hemisphere being negligible.}
\end{figure}

The above-mentioned distinction between various continuous driving forms is further supported by the phase-space structure of the selected Floquet modes.
The Husimi distributions of the four modes S, T, U and V, which were marked in Fig.\,\ref{fig:VD}, are displayed in Fig.\,\ref{fig:VD_Husimi}.
We see that the modes T and V corresponding to the Cos2 and Tent drivings, respectively, show essentially the same locations of the elliptic periodic points as the Delta driving in panel D of Fig.\,\ref{fig:HD_PL}.
In contrast, modes S and U corresponding, respectively, to the Cos1 and Saw drivings show different locations (the elliptic periodic points placed closer to the equator of the Bloch sphere), which explains the opposite direction of the resonance distortion of the lattice with respect to the Delta driving.

It should be stressed that the similarity of the Peres lattices for various types of driving is preserved only for moderate values of the driving strength $\eta$.
We have verified (the results not shown here) that if $\eta$ increases, the regular lattices of the weakly perturbed system decay in different driving-specific ways.
Small regions of regularity around ${\eta=n\pi}$, present for the Delta driving (Sect.\,\ref{sec:chaos}), do not appear for the other driving forms.

\subsection{Floquet Hamiltonian approach
\label{sec:EH}}

For continuously driven systems, the construction of the Floquet Hamiltonian $\widehat{G}$ in Eq.\,\eqref{eq:FloquetHamiltonian} is less straightforward than for kicked systems (Sect.\,\ref{sec:BCH}).
Several methods exist that can produce different forms of $\widehat{G}$ on different levels of approximation \cite{Honsa2022, Bandyopadhyay2015, Goldman2014, Bastidas2014b, Rahav2003, Blanes2009}.
Here, we use the method described in Ref.\,\cite{Rahav2003}, which simultaneously constructs the Floquet Hamiltonian $\widehat{G}$ and the kick operator $\widehat{K}(t)$. 
Its advantage is a simple form of $\widehat{G}$, the disadvantage is that usually $\widehat{K}(t)\neq 0\,\forall t$. 
Technical details can be found in Refs.\,\cite{Rahav2003, Goldman2014} and examples of application in Refs.\,\cite{Sharma2018, Bandyopadhyay2015}.
Roughly, the method makes use of the Fourier series of the full Hamiltonian \eqref{eq:DrivenLH},
\begin{eqnarray}
    \widehat{H}(t) = \widehat{H}_0 + \frac{\eta}{T}\widehat{H}'\sum_{n = 1}^{+\infty}\left( g_n e^{\mathrm{i} 2 \pi n t /T} \!+\! g^*_n e^{-\mathrm{i} 2 \pi n t /T} \right),
    \label{eq:Hperexp}
\end{eqnarray}
where $g_n$ satisfying ${|g_n|\sim 1}$ are Fourier expansion coefficients of the function $Tg(t)$, cf.\,Eq.\,\eqref{eq:expdelta}.
Individual $\propto T^n$ terms of the Dyson expansion of the evolution operator over one period $T$ are compared with the corresponding terms obtained by the expansion of the exponentials on the right-hand side of Eq.\,\eqref{eq:FloquetHamiltonian}.
Because of the $\eta/T$ prefactor in Eq.\,\eqref{eq:Hperexp}, the resulting series has the form given in Eq.\,\eqref{eq:expansions}.

Here we show the series truncated at ${N = 2}$ for the drivings from Eqs.\,(\ref{eq:g_Cos1}) and (\ref{eq:g_Cos2}).
For the Cos1 driving we obtain:
\begin{eqnarray}
	\widehat{G}^{(2)} &=& \widehat{J}_z - \frac{\kappa}{2 j} \widehat{J}_x^2 + \frac{1}{4\pi^2} \cdot \frac{\kappa \eta^2}{j} \left( \widehat{J}_x^2 - \widehat{J}_y^2\right),
    \label{eq:G_Cos1}\\
	\widehat{K}(t)^{(2)} &=& -\frac{\eta}{\pi } \sin \left( \frac{2 \pi t}{T} \right) \widehat{J}_z + 
    \label{eq:K_Cos1}\\ 
    &+& \frac{1}{4 \pi ^ 2} \cdot \frac{\kappa \eta T}{j}  \cos\left( \frac{2\pi t}{T}\right) \left( \widehat{J}_x \widehat{J}_y+ \widehat{J}_y \widehat{J}_x \right). 
    \nonumber
\end{eqnarray}
For the Cos2 driving we get:
\begin{eqnarray}
	\widehat{G}^{(2)} &=& \widehat{J}_z -\frac{\kappa}{2 j} \widehat{J}_x^2 + \frac{5}{16\pi^2} \cdot \frac{\kappa \eta^2}{j} \left( \widehat{J}_x^2 - \widehat{J}_y^2\right),
    \label{eq:G_Cos2}\\
	\widehat{K}(t)^{(2)} &=& \frac{\eta}{2\pi } \left[ -2 \sin \frac{2 \pi t}{T} + \sin\frac{4\pi t}{T} \right] \widehat{J}_z + 
    \label{eq:K_Cos2}\\  
    &+& \frac{\kappa \eta T}{4 \pi ^ 2 j} \left[ \cos \frac{2\pi t}{T} - \frac{1}{4} \cos \frac{4\pi t}{T} \right] \left( \widehat{J}_x \widehat{J}_y+ \widehat{J}_y \widehat{J}_x \right).
	\nonumber
\end{eqnarray}
The second-order approximations of the Floquet Hamiltonians in Eqs.\,\eqref{eq:G_Cos1} and \eqref{eq:G_Cos2} can be compared with the corresponding formula \eqref{eq:G2BCH} for the kicked system.

\begin{figure}
	\includegraphics[width=\columnwidth]{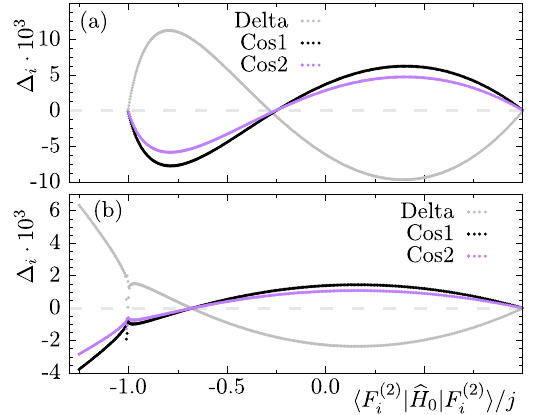}
	\caption{\label{fig:EH}
    Deviations \eqref{eq:deviace} of the $\widehat{H}_0\times\widehat{J}_z$ Peres lattice for the second-order truncated Floquet modes of the Cos1 and Cos2 drivings from the corresponding lattice of the stationary system. The truncated modes for both drivings result from Eqs.\,(\ref{eq:G_Cos1})-(\ref{eq:K_Cos1}) and (\ref{eq:G_Cos2})-(\ref{eq:K_Cos2}). We also show the deviations for the Delta driving with the second-order truncated Floquet modes resulting from Eq.\,\eqref{eq:G2BCH}. The parameters are: (a) ${\kappa = 0.7}, {T = 7.15}, {\eta = 0.2}, {j = 200}$ (upper panel) and (b) ${\kappa = 2}, {T = 5}, {\eta = 0.01}, {j = 200}$ (lower panel).}
\end{figure}

To compare the Floquet modes resulting from the truncated Floquet Hamiltonian approach with the exact ones, we have to use the whole formula \eqref{eq:FloquetHamiltonian} with the consistently truncated forms of $\widehat{G}^{(N)}$ and $\widehat{K}(t)^{(N)}$ for the evolution over one period.
Here we approximate the Floquet evolution operator by 
\begin{equation}
 \widehat{U}(T, 0)^{(2)} = \mathrm{e}^{-\mathrm{i} \widehat{K}(0)^{(2)}} \mathrm{e}^{-\mathrm{i} \widehat{G}^{(2)}T} \mathrm{e}^{+\mathrm{i} \widehat{K}(0)^{(2)}}   
\end{equation}
and calculate the corresponding truncated Floquet modes $\ket{F^{(2)}_i}$.
Again, as in the case of the kicked system (see the left panels of Fig.\,\ref{fig:BCH}), it turns out that Peres lattices of the $\ket{F^{(2)}_i}$ modes deduced from Eqs.\,(\ref{eq:G_Cos1})-(\ref{eq:K_Cos1}) and (\ref{eq:G_Cos2})-(\ref{eq:K_Cos2}) are very similar to the lattices of the stationary system, rather than to the exact lattices of the respective driven systems.
This is illustrated in Fig.\,\ref{fig:EH}, where we show the quantity 
\begin{eqnarray}
	\Delta_i = \frac{\braket{ F^{(2)}_i | \widehat{J}_z| F^{(2)}_i }}{j} - J_z \! \biggl(\mathcal{E} = \frac{\braket{ F^{(2)}_i | \widehat{H}_0| F^{(2)}_i }}{j}\biggr),
    \label{eq:deviace}
\end{eqnarray}
which captures the deviation of the expectation value of~$\widehat{J}_z$ in the truncated Floquet mode from the interpolated expectation value in the stationary lattice, which is given by the function $J_z(\mathcal{E})$ from Eq.\,\eqref{eq:inter}. 
We plot $\Delta_i$ for each truncated Floquet mode as a function of the expectation value of $\widehat{H}_0$ for ${j = 200}$.
We also show in Fig.\,\ref{fig:EH} the analogous ${N=2}$ deviations for the Delta driven system, where, however, we use the different choice of initial time $t_0$ described in Sect.\,\ref{sec:BCH} (the choice compatible with the present case would modify the shape but not the overall size of the $\Delta_i$ dependence). 
Note that the similarity measure \eqref{eq:measure} between the respective Peres lattices can be written as $\mathcal{N}=\sum_i\Delta_i^2/(2j+1)$. 

The negligible values of $\Delta_i$ read in Fig.\,\ref{fig:EH} indicate that the second-order approximations of the Floquet modes yield Peres lattices that are almost identical to those produced by the stationary Hamiltonian.
We do not observe any trace of the resonances that appear in the exact lattices of driven systems.
This is a consequence of small correction terms that extend the unperturbed Hamiltonian~$\widehat{H}_0$ in expressions (\ref{eq:G_Cos1}) and (\ref{eq:G_Cos2}) of $\widehat{G}^{(2)}$.
The prefactors in front of the correction terms are approximately equal to $4 \cdot 10^{-6}$ for the first set of parameters and $3 \cdot 10^{-8}$ for the second set of parameters. 
The operator norm (defined as the largest singular value) assigned to the ${N=2}$ corrections takes the following values: (a) 0.14 and 0.18 for the Cos1 and Cos2 drivings, respectively, with the first set of parameters, and (b) 0.001 for both types of driving with the second set of parameters.
Therefore, the present approximation of the Floquet Hamiltonian does not depart significantly from $\widehat{H}_0$, and it also yields very similar Peres lattices.
We stress that the computation of the higher-order terms would become extensively difficult, while divergent behavior similar to that for the BCH formula (Sect.\,\ref{sec:BCH}) may be anticipated.

\section{Conclusions\label{sec:conc}}

In this paper, we explore the applicability of Peres lattices in periodically driven systems.
The Peres lattice, originally derived from the expectation values of two selected observables in the eigenstates of a stationary Hamiltonian, is redefined using the Floquet modes of the periodically driven system. 
We confirm the same role of Peres lattices in the driven systems as in stationary systems, namely, their ability to provide a holistic diagnostic tool for detecting perturbations of dynamics in the studied system with respect to a reference dynamics.
As a byproduct, we disclose several particular features of the system used as a~testing ground, which was the fully connected set of qubits subject to $\delta$-kicked or continuous types of periodic driving.  

We test various types of Peres lattice (various Peres observables) and show that the most important features are visible in \lq\lq almost any\rq\rq\ (randomly chosen) lattice  (of course with case-specific exceptions) with an arbitrary choice of the initial time.
In particular, we demonstrate how the Peres lattices reflect the following attributes of the periodic driving:
\begin{itemize}
    \item Appearance of resonances: 
    The dynamics near some classical orbits of the unperturbed system that happen to be in resonance with the external driving is most strongly affected and leads to the onset of chaos. 
    This is seen as a distortion of the Peres lattice relative to the stationary system, which appears already for relatively weak driving strengths. 
    As the size of the system increases, smaller resonances can be distinguished in the Peres lattice.
    \item Depiction of the transition to chaos: 
    With an increasing strength of the driving, the distortions of the Peres lattice leak out from the resonance regions and affect larger parts of the lattice. 
    Even for a system with one degree of freedom, the lattice loses its one-dimensional chain-like nature and acquires a two-dimensional character. 
    Once chaos starts to dominate, ergodicity shrinks the entire lattice into a small interval of expectation values. 
    Nevertheless, we demonstrate that in some cases full regularity can reappear periodically even for very large driving strengths.
    \item Signatures of dynamical criticality: 
    Critical dynamics near classical stationary points, which leads to excited-state phase transitions in the spectra of stationary systems, can survive even after the application of periodic driving (cf.\,Refs.\,\cite{Bandyopadhyay2015, Bastidas2014b, Bastidas2014a}). 
    This is distinguished in cusps or other singular shapes in the Peres lattices and confirmed by the analysis of the phase-space images (Husimi distributions) of the corresponding Floquet modes. 
    While the critical phenomena have been established independently in the literature, the Peres lattice provides a strong indication of their presence and offers a direct way of their visualization.
    Peres lattices can also reflect spontaneous breaking of symmetry (here the parity), which is sometimes associated with the excited-state criticalities.
    \item Insight into the type of driving: 
    Peres lattices represent a powerful tool for quick identification (and localization) of the dynamical differences (if any) between periodically driven systems with various driving functions and/or driving operators. 
    \item Approximations by truncated Floquet Hamiltonians: 
    Peres lattices enable us to quickly test the quality of approximations of the driven dynamics by means of truncated expansions of the Floquet Hamiltonian and the corresponding kick operator. 
    In particular, the performance of such approximations was shown to be very poor in our test system.
\end{itemize}

In conclusion, our work shows that the method of Peres lattices represents a~comprehensible tool for analyzing various properties of periodically driven quantum systems.
This tool is relatively easy and straightforward to access from the computational point of view, yet it provides very compact information on all Floquet modes in a single picture. 
We believe that the method can facilitate future theoretical studies of periodically driven systems, possibly in connection with quantum simulation and sensing applications.

\begin{figure*}
	\includegraphics[width=\textwidth]{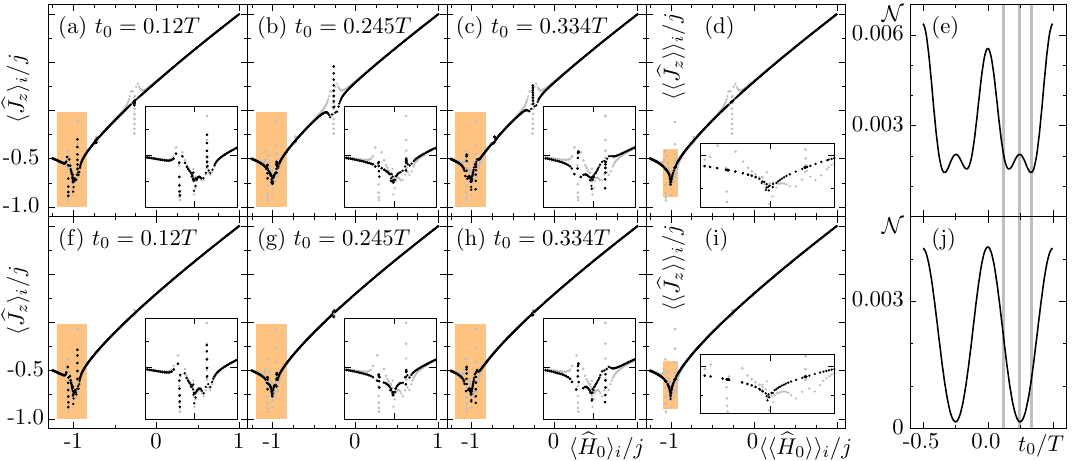}
	\caption{\label{fig:T_0}
    Peres lattices for various initial times $t_0$ of the driving [panels (a)--(c) and (f)--(h)], the lattices averaged over $t_0$ [panels (d) and (i)], and the dependence of the dissimilarity measure \eqref{eq:measure} on $t_0$ [panels (e) and (j)]. Upper [(a)--(e)] and lower [(f)--(j)] rows correspond to the Delta and Cos1 drivings, respectively. The lattices for $t_0 = 0$ are shown in light gray. The insets depict zoomed-in areas. The vertical lines in panels (e) and (j) indicate the $t_0$ values in panels (a)--(c) and (f)--(h). The parameters are $\kappa = 2, T = 5$, and $\eta = 0.01$.}
\end{figure*}

\begin{acknowledgments}
    The authors thank P. Stránský for valuable discussions.
	L.H. and P.C. acknowledge financial support from the Czech Science Foundation under project no.\,25-16056S. 
    L.H, J.S., and J.N. acknowledge financial support from the Charles University Grant Agency under project no.\,215323 and from the Charles University Research Centre of Excellence UNCE/24/SCI/016. 
\end{acknowledgments}

\section*{Data availability}
 The data and code that support the findings of this article are publicly available \cite{PeresLatticesZenodo2026}.

\appendix

\section{Dependence on the initial time}\label{sec:App_T_0}

In this appendix, we focus on the dependence of the Peres lattice on the initial time of the driving. 
The expectation values for Floquet modes with two choices $t_0$ and $t_0'$ of the initial time are related by a unitary transformation of the corresponding observable:
\begin{eqnarray}
\!\!\langle F_i (t_0)| \widehat{O} | F_i(t_0)\rangle =\langle F_i (t_0')| \underbrace{\widehat{U}^\dagger(t_0, t_0') \widehat{O} \widehat{U}(t_0, t_0')}_{\widehat{O}'} | F_i(t_0')\rangle \nonumber.\\
~
\end{eqnarray}
In Fig.\,\ref{fig:T_0} we show Peres lattices for the Delta and Cos1 drivings\,---\,see panels (a)--(c) and (f)--(h), respectively\,---\,with various values of $t_0$.
We see that the resonances are in the same places, but have different shapes. 
Depending on $t_0$, the deviations from the unperturbed lattice at the resonance positions take both positive and negative values. 
In panels (d) and (i) we show the lattices obtained by averaging of $\langle\widehat{O}\rangle_i$ over $t_0$:
\begin{eqnarray}
\langle\langle\widehat{O}\rangle\rangle_i = \frac{1}{T}\int_{-T/2}^{+T/2} \langle F_i (t_0)| \widehat{O} | F_i(t_0)\rangle d t_0.
\end{eqnarray}
In this case, significant deviations from the unperturbed lattice disappear and the traces of resonances are only visible on closer inspection (see the insets).

In panels (e) and (j) of Fig.\,\ref{fig:T_0}, we show the $t_0$-dependence of the dissimilarity $\mathcal{N}$ of the actual lattice from the unperturbed one, see Eq.\,\eqref{eq:measure}.
The minima in these plots correspond to the cases in which the impact of resonances on the lattice is minimal.
However, even in these cases the value of $\mathcal{N}$ is not zero, indicating that the resonances are not generically suppressed altogether; see panels (c) and (g).

\section{Convergence of the evolution-operator expansions\label{sec:App_Convergence}}

\begin{figure*}
	\includegraphics[width=\textwidth]{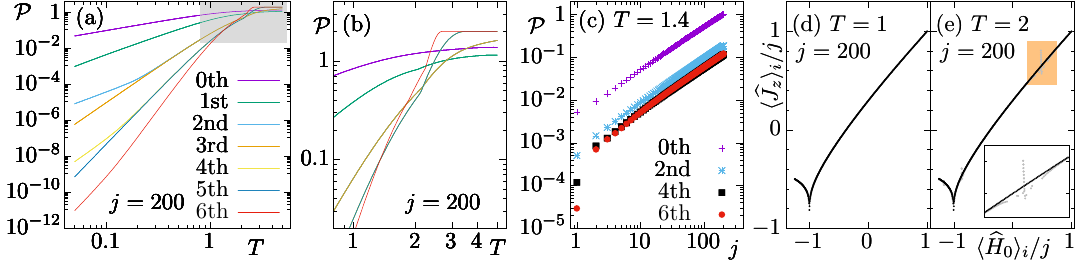}
	\caption{\label{fig:Convergence_BCH}The quality of the BCH approximation for the kicked system. We draw the error measure \eqref{eq:PropNorm} for various orders of the expansion as a function of the period $T$ [panels (a) and (b) (zoomed area)] and the size parameter $j$ [panel (c)]. We also show examples of Peres lattices for the 6th-order BCH approximation [panels (d) and (e)] in black, with the precise counterparts in gray. The parameters are $\kappa = 2, \eta = 0.01$.
    }
\end{figure*}

\begin{figure}
	\includegraphics[width=\columnwidth]{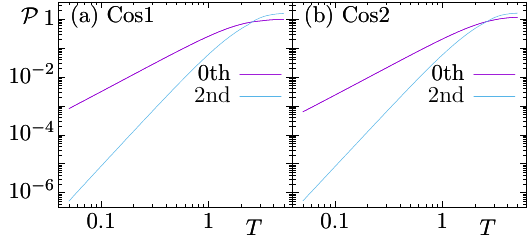}
	\caption{\label{fig:Convergence_Cos12}Error measure \eqref{eq:PropNorm} for the 0th- and 2nd-order expansion as a function of the period $T$ for the continuous Cos1 and Cos2 drivings. 
    The parameters are $\kappa = 2$, $\eta = 0.01$, and $j = 200$.}
\end{figure}

In this appendix, we look more closely at the convergence of the expansions involved in the Floquet-Hamiltonian approach. 
We use an error measure defined as the largest singular value of the difference between the exact one-period evolution operator and its approximate counterpart,
\begin{eqnarray}
    \mathcal{P} = \|\widehat{U}(t_1, t_0)- \mathrm{e}^{-\mathrm{i}\widehat{K} (t_1)^{(N)}} 
    \mathrm{e}^{-\mathrm{i}(t_1 - t_0) \widehat{G}^{(N)}} 
    \mathrm{e}^{\mathrm{i}\widehat{K}(t_0)^{(N)}}\|.\nonumber\\
    ~\label{eq:PropNorm}
\end{eqnarray}
We stress that our results are independent of the choice of the logarithm branch in the evaluation of $\widehat{G}^{(N)}$ and $\widehat{K}(t)^{(N)}$, since we use only quantities that can be derived from the full exponentials.

We begin with the kicked system and BCH formula discussed in Sec.\,\ref{sec:BCH}. 
The choice $t_0 = T/2-\epsilon$ and $t_1 = 3T/2-\epsilon$ implies that $\widehat{K}(t_0)^{(N)} = \widehat{K}(t_1)^{(N)} = 0$. 
We recall that the convergence of the expansions \eqref{eq:expansions} is anticipated when both $\eta$ and $T$ are sufficiently small.
Indeed, Fig.\,\ref{fig:Convergence_BCH} shows that for small driving period $T$ the higher-order terms improve the accuracy of the expansion, see panels (a) and (b).
However, in the convergent cases the Peres lattice is virtually identical with the unperturbed one, see panel (d), and when the resonances occur, panel (e), the expansion is unable to capture them.

Figure \ref{fig:Convergence_BCH}(c) shows the dependence of the error measure \eqref{eq:PropNorm} on the size parameter $j$. 
Since the norms of the operators in the exponentials of Eq.\,\eqref{eq:BCHdef} decrease with decreasing $j$, the error measure decreases as well.
Interestingly, for the chosen set of parameters, the 4th-order approximation yields a smaller error than the 6th-order approximation for $j\gtrsim 16$, whereas the opposite is true for $j<16$.
This indicates that the truncated series can be relatively accurate even when the full series does not converge \cite{Kuwahara2016}.

Next, we discuss the expansions for the continuous Cos1 and Cos2 driving protocols from subsection \ref{sec:EH}. 
In Fig.\,\ref{fig:Convergence_Cos12}, we plot the dependence of the error measure \eqref{eq:PropNorm} on the period $T$ for 0th-order approximation defined by $\widehat{U}(T, 0)\approx e^ {-\mathrm{i}T\widehat{H}_0}$ and the 2nd-order approximation from Eqs.\,\eqref{eq:G_Cos1}--\eqref{eq:K_Cos2}. 
Note that for $t_0 = 0$, the 1st-order approximation is the same as the 0th-order one. 
We observe the same trend as for the kicked system: for larger values of $T$, the 0th-order approximation yields better results than the 2nd-order one, while the opposite is true for smaller $T$.

\section{On the numerical precision}\label{sec:App_num}

In this appendix, we comment on the numerical techniques and precision of our calculations.
The kicked system has an advantage in a simple form of the one-period evolution operator \eqref{eq:UKT}. 
Thus, the calculation of the evolution operator comprises three matrix exponentials and their multiplication. 
For continuous drivings, the evolution is computed via Hamiltonians that are constant during sufficiently short equal-length intervals.
In all calculations, we use 256 intervals and verify that increasing this number has a negligible effect on the lattice. 
For instance, doubling the number of intervals to 512 produces shifts of the lattice points in Fig.\,\ref{fig:VD} of order $10^{-5}$ relative to the overall lattice size.

Most calculations were performed in double precision using a standard library for matrix exponentiation and diagonalization.
We checked that numerical errors are negligible.
The results presented in Fig.\,\ref{fig:BCH} for the BCH formula were calculated in octuple precision to rule out significant numerical errors. 
We checked that quadruple precision for the 8th-order expansion and double precision for the 6th- and lower-order expansion would be sufficient. 

A challenging problem is accurately evaluating the properties of the nearly-degenerate parity doublets that occur (for parity-conserving drivings) in the parity-breaking phase and possibly in certain resonance regions. A straightforward solution is to diagonalize the evolution operator in separate positive- and negative-parity subspaces. However, since the lattice observables used above yield virtually the same expectation values for both members of the parity doublet, the separate diagonalization is not even necessary. We checked that the errors for parity doublets in our calculations (in the full space) are negligible, e.g., on the level of $10^{-11}$ of the overall lattice size in Fig.\,\ref{fig:big}.

\bibliography{Database_A}

\end{document}